\documentclass[table]{gtech}
\PassOptionsToPackage{table, usenames, dvipsnames}{xcolor}
\usepackage{xcolor}
\usepackage{amssymb}
\usepackage{multirow}
\usepackage{bigdelim}
\usepackage{longtable}
\usepackage{tabularray}
\usepackage{wrapfig}
\usepackage[most]{tcolorbox}
\usepackage{url}
\usepackage{float}
\usepackage{datatool}
\usepackage{enumitem}
\usepackage{subcaption} %
\usepackage[justification=centering]{caption}

\usepackage[most]{tcolorbox}
\usepackage{xcolor}

\newtcolorbox{graybox}{
  notitle,
  rounded corners,
  colframe=darkgray,
  colback=white,
  boxrule=0.5mm,
  arc=1mm,
  left=2pt,
  right=2pt,
  top=4pt,
  bottom=4pt,
  after skip=5pt,
  before skip=5pt
}

\RequirePackage{tgpagella} %
\RequirePackage{mathpazo}  %
\RequirePackage{inconsolata} %
\usepackage{makecell}
  \usepackage{booktabs}
  \usepackage{tabularx}
  \usepackage{array}
\usepackage{booktabs} %
\usepackage{adjustbox}
\usepackage{pgfplots}
\pgfplotsset{compat=1.18}
\usepackage[utf8]{inputenc} %
\usepackage[T1]{fontenc}    %
\usepackage{hyperref}       %
\usepackage{url}            %
\usepackage{booktabs}       %
\usepackage{amsfonts}       %
\usepackage{nicefrac}       %
\usepackage{microtype}      %
\usepackage{xcolor}         %
\usepackage{xspace}
\usepackage{amsthm}   %
\usepackage{amsmath}  %
\usepackage{amssymb}  %
\usepackage{bm}       %
\usepackage{algorithm}
\usepackage{algpseudocode}
\usepackage[normalem]{ulem}
\theoremstyle{definition}

\usepackage{booktabs}
\usepackage{tabularx}
\newtcolorbox{bluebox}{
  colback=blue!5,
  boxrule=0.5mm,
  arc=1mm,
  left=2pt,
  right=2pt,
  top=4pt,
  bottom=4pt,
  after skip=5pt,
  before skip=5pt
}

\renewcommand{\title}[1]{\newcommand{\titlelist}{{\huge\fontfamily{optimistic}\selectfont #1}}}

\newcommand{\ignore}[1]{}

\usepackage{arydshln}
\definecolor{CQColor}{rgb}{0.0,0.0,1.0} %

\usepackage{colortbl}
\usepackage{amssymb}
\usepackage{pifont}
\usepackage{booktabs,multirow}
\usepackage{makecell}
\usepackage{tabulary}
\usepackage{fontawesome5}
\usepackage{bbding}
\usepackage{multicol}

\newlength\savewidth

\title{Towards a Densing Law for User Representation Learning at Billion-Scale Capacity}

\author[1,*]{Bin Dou}
\author[1,2,*]{Junru Zhang}
\author[1,2,*]{Zhaoyi Yuan}
\author[1]{Wuliang Huang}
\author[1]{Letian Gong}
\author[1,\dag]{Baokun Wang}
\author[2,\dag]{Huan Li}
\author[1]{Yu Cheng}
\author[1]{Weiqiang Wang}

\affiliation[1]{DeepFind Team, Ant Group}
\affiliation[2]{Zhejiang University}

\contribution[*]{\footnotesize Equal Contribution}
\contribution[\dag]{\footnotesize Corresponding author
\email{yike.wbk@antgroup.com},
\email{lihuan.cs@zju.edu.cn}
}

\abstract{
User representation learning in real-world industrial scenarios is commonly scaled by increasing
user amount, behavioral sequence length, and model size. However,
existing methods suffer from two issues: (i) \textit{Bottleneck for raw data scaling at billion-scale capacity}, as performance gains diminish with larger-scale raw text user behavioral input which can be alleviated by tokenization. (ii) \textit{Lack of quantitative analysis of how tokenization configurations should scale with data size}.
In this paper, we propose \emph{User Behavioral Densing Law} for 
characterizing the quantitative relationship between data scale and the minimum sufficient tokenization capacity.
Firstly, we conduct pilot study on raw \& tokenized scaling comparison on billion-scale Alipay dataset, revealing the raw data scaling bottleneck and breakthrough via tokenization.
To derive the scaling pattern governing the minimum sufficient tokenization configuration at different data scales,
theoretical and experimental analyses are employed to summarize the quantitative scaling pattern.
In addition, we propose a new tokenization method named ALGN inspired by the proposed Densing Law.
Experimental evaluation over different data sources, tokenization methods and downstream tasks proves the generalizability and reliability of the Densing Law, which provides the guidance for configuration selection in large -scale user representation learning. Also ALGN surpasses existing baselines on both performance and efficiency.
}

\begin{document}
  \maketitle

\section{Introduction}
\label{sec:intro}

The remarkable success of deep representation learning has fundamentally transformed modern personalized platforms, with downstream performance consistently improving as user behavioral sequences and model parameters scale~\citep{kang2018self, sun2019bert4rec, pi2019practice}. Today, unified user embeddings extracted from raw behavioral logs serve as a shared foundation across recommendation, CTR/CVR prediction, retrieval, and natural language user targeting~\citep{xiao2025mars, zhang2021deep}. Figure~\ref{fig:teaser} (a) sketches this paradigm. However, scaling along practical axes, including more users, longer historical windows, and larger encoders~\citep{ardalani2022understanding, guo2024scaling, zhang2024scaling}, presents significant challenges. User behavior logs are heavily shaped by daily routines and repeated habits, meaning data volume grows rapidly while novel, task-relevant information grows much more slowly. This discrepancy introduces a critical  question:

\begin{figure}[!ht]
  \centering
  \includegraphics[width=\linewidth]{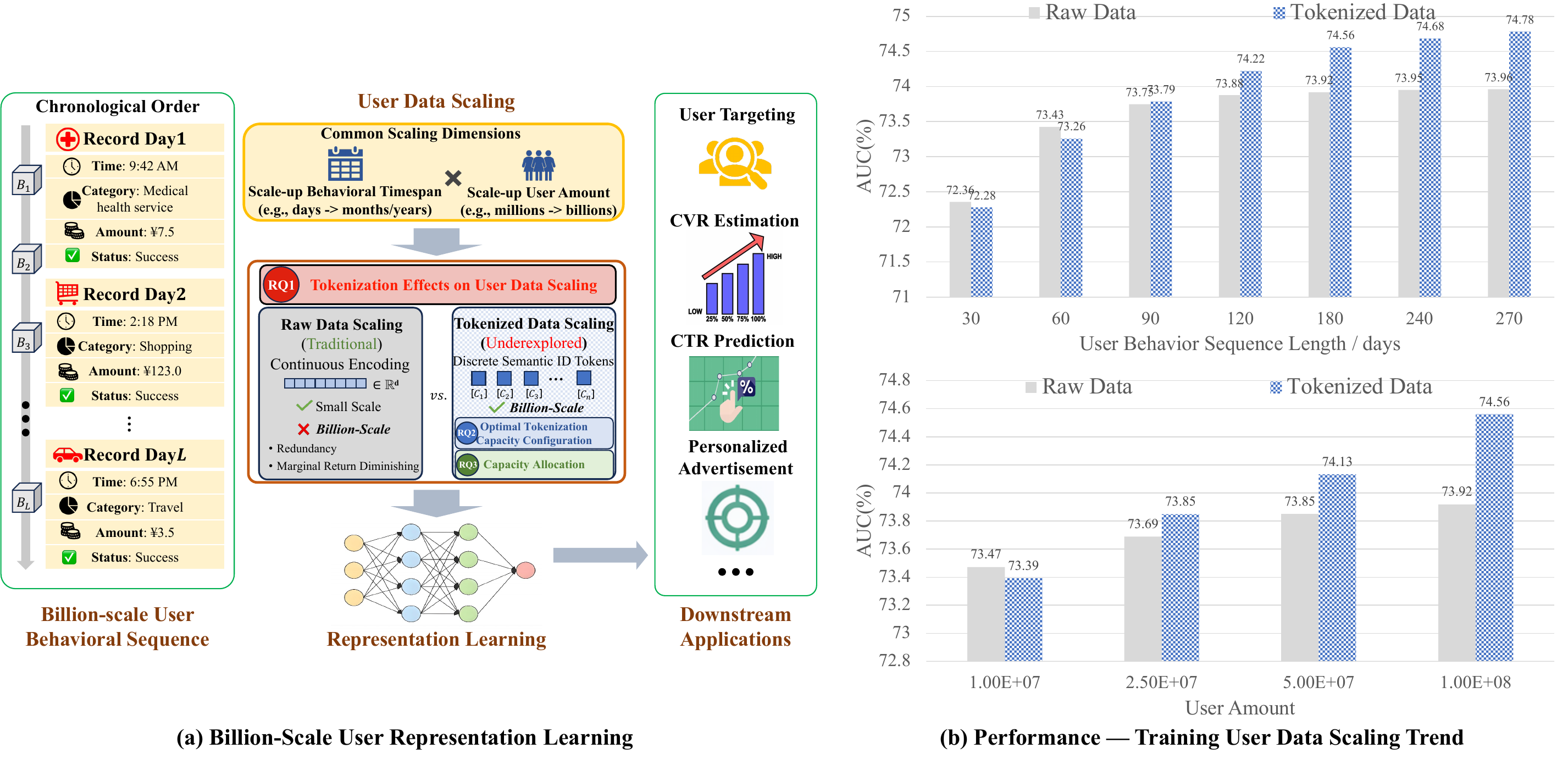}
  \caption{
      (a) User representation learning pipeline. Once trained on user behavioral sequences, a representation model can support various downstream applications, including text-based retrieval for user targeting, U2U retrieval for recommendation system and classification for advertising and risk control scenarios.
      In industrial settings with billion-scale users, when scaling up input data volume, commonly via increasing behavioral timespan and user amount, models directly consume raw data (e.g., raw user behavioral text description), will suffer from performance saturation beyond practical scaling thresholds while tokenization will alleviate the limitation and enable sustained performance gains, as shown in (b).
  }
  \label{fig:teaser}
\end{figure}

\begin{bluebox}
\begin{center}
\emph{Can we identify the key properties that make scaled models effective, thus improving representation quality without simply processing a greater volume of raw behavioral events?}
\end{center}

\end{bluebox}

While existing industrial systems rely on scaling raw sequence data, the empirical manifestation of behavioral redundancy and its relationship to downstream performance remain underexplored.
In this work, we provide a comprehensive empirical analysis of what we term the \textit{raw behavioral scaling wall}, using large-scale real-world Alipay PayBill data to ensure practical relevance and ecological validity.

\begin{graybox}
\textbf{Definition 1.1.}
We define the \textit{raw behavioral scaling wall} as the phenomenon where incrementally scaling raw behavioral sequence length, user population, or model capacity yields strictly diminishing downstream returns, culminating in a \textit{loss and quality dissociation} where pre-training loss decreases without improving downstream accuracy.

\end{graybox}

Through systematic evaluations across multiple scaling axes, we demonstrate that industrial user representation learning exhibits severe redundancy saturation. Extending observation windows or user bases primarily introduces repetitive, low-information events. Most notably, when scaling a user encoder from 0.2B to 0.4B parameters, we observe that the larger model merely fits redundant behavioral details rather than extracting additional task-relevant signal, leaving downstream representation accuracy almost unchanged.

\begin{graybox}
\textbf{Definition 1.2.}
We define \textit{behavioral densing} as the process of transforming long raw histories into compact representations that preserve task-relevant distinctions while suppressing redundancy, thereby optimizing the amount of downstream-relevant signal each effective input unit carries.

\end{graybox}

This density-oriented perspective shows a fundamental insight: information density, rather than model capacity alone, is a key bottleneck limiting the expressivity of user encoders. We hypothesize that moving user representations toward a better performance-cost frontier, governed by the \textit{behavioral densing law}, can recover the gains that raw scaling fails to provide. This law rests on three principles: \textit{Bounded Raw Gains}, where marginal utility drops as redundancy dominates; \textit{Density Driven Scaling}, where denser representations shift the saturation frontier; and \textit{Minimal Sufficient Capacity}, where optimal representations allocate only enough capacity to preserve necessary task-relevant information.

To test this hypothesis, we instantiate behavioral densing using residual quantized behavioral tokenization via RQ-VAE. This acts as a practical density operator, mapping repetitive event histories into compact sequences of discrete codes. Furthermore, recognizing that behavioral information is not uniformly distributed, we propose an \textit{entropy-based variable length tokenization} strategy. By allocating more discrete codes to high-entropy segments and fewer to predictable routines, this mechanism explicitly adapts to behavioral complexity and approaches minimal sufficient representation capacity.

Our empirical evaluations demonstrate that under matched token and compute budgets, the tokenized representations overcome the raw scaling wall. The proposed framework yields consistent performance gains and improved encoder utilization on large-scale Alipay PayBill production data.

The key contributions of this work are:

\begin{enumerate}

\item \textbf{Scaling diagnosis and tokenization effects.}
We identify diminishing marginal returns when scaling raw behavioral data, and reveal that tokenized data deliver more sustained gains thereby mitigating performance saturation at larger scales. Together with evidence of a model-capacity bottleneck, these findings indicate that data information density, rather than model parameter number alone, is a primary constraint on industrial user representation learning, especially at billion-scale data capacity.

\item \textbf{Predictive tokenization configuration recipe.}
We formulate the optimal tokenization capacity configuration trajectory corresponding to input data volume as the \textit{Behavioral Densing Law} for user representation learning, enabling required capacity configurations to be estimated from behavioral data scale and lightweight statistics after tokenizer-method calibration.

\item \textbf{Adaptive tokenization allocation.}
We propose \textit{Adaptive Length Gated Network}, a novel approach which adaptively allocates tokenization capacity via variable residual depth across user samples according to the quantization residual and expression uncertainty, enabling more effective allocation of the discrete representation space and exhibiting advantages as the input data scales up from the \textit{Densing Law} perspective.

\end{enumerate}

\section{Related Work}
\label{sec:related}

Existing studies related to this work can be organized into three
directions: scaling raw behavioral data, constructing tokenized
behavioral representations, and adapting token usage across inputs.
Table~\ref{tab:related_work} summarizes their coverage of the dimensions
considered in this study.

\begin{table}[ht]
    \centering
    \caption{Comparison of related work across the dimensions considered
    in this study.}
    \label{tab:related_work}

    \scriptsize
    \setlength{\tabcolsep}{2.2pt}
    \renewcommand{\arraystretch}{1.15}

    \begin{tabularx}{\columnwidth}{
        @{}
        p{0.40\columnwidth}
        *{4}{>{\centering\arraybackslash}X}
        @{}
    }
        \toprule

        \textbf{Research direction}
        &
        \shortstack{\textbf{Raw}\\\textbf{Data}}
        &
        \shortstack{\textbf{Tokenized}\\\textbf{Data}}
        &
        \shortstack{\textbf{Capacity}\\\textbf{Configuration}}
        &
        \shortstack{\textbf{Capacity}\\\textbf{Allocation}}
        \\

        \midrule

        User data scaling
        \citep{ardalani2022understanding,
        shin2021scaling,zhang2023scaling,
        zivic2024scaling,guo2024scaling,
        zhai2024actions,shen2025plaw,
        zhang2026length}
        &
        $\checkmark$
        &
        $\times$
        &
        $\times$
        &
        $\times$
        \\

        Behavioral tokenization
        \citep{feng2026behavior,
        liu2024discrete,zhu2024cost,
        liu2024learning,hou2025actionpiece}
        &
        $\times$
        &
        $\checkmark$
        &
        $\times$
        &
        $\times$
        \\

        Residual quantization based tokenization
        \citep{vqvae,
        lee2022autoregressive,
        rajput2023recommender,
        he2025learning}
        &
        $\times$
        &
        $\checkmark$
        &
        $\times$
        &
        $\times$
        \\

        Adaptive quantization
        \citep{huijben2024residual,
        seo2024rate,chae2025variable,
        kusupati2022matryoshka}
        &
        $\times$
        &
        $\checkmark$
        &
        $\times$
        &
        $\checkmark$
        \\

        \midrule

        \textbf{Ours}
        &
        $\checkmark$
        &
        $\checkmark$
        &
        $\checkmark$
        &
        $\checkmark$
        \\

        \bottomrule
    \end{tabularx}
\end{table}

\paragraph{Raw behavioral data scaling.}
Large scale user modeling studies examine how behavioral data, model
capacity, and computation affect downstream utility
\citep{ardalani2022understanding,shin2021scaling,
zhang2023scaling,zivic2024scaling,guo2024scaling,
zhai2024actions}. Long history methods improve the utilization of
extended behavioral sequences through interest extraction, retrieval,
and long sequence architectures
\citep{zhou2019deep,li2019multi,cen2020controllable,
pi2020search,ren2025longretriever,cao2022sampling,
si2024twin,zhou2024encode,chai2025longer}.
However, additional records may contain repetitive or weakly informative
signals, resulting in diminishing marginal gains
\citep{shen2025plaw,zhang2026length}. These studies primarily
characterize the scaling behavior of raw behavioral data.

\paragraph{Tokenized behavioral data.}
Behavioral tokenization converts raw user data into compact discrete
representations. Existing methods learn reusable behavioral units or
vocabularies for user understanding
\citep{feng2026behavior,liu2024discrete,zhu2024cost,
liu2024learning,hou2025actionpiece,rqkmeans,sarq}.
VQ-VAE~\citep{vqvae} learns discrete latent codes through vector
quantization, while RQ-VAE~\citep{lee2022autoregressive} recursively
quantizes residual information using multiple codebooks. Based on this
formulation, TIGER constructs semantic identifiers for generative
retrieval \citep{rajput2023recommender}, and U$^{2}$QT introduces
multi view quantization for compact user tokens
\citep{he2025learning}. These methods generally use fixed tokenization
configurations at selected data scales.

\paragraph{Capacity configuration and adaptive tokenization.}
Codebook size, residual depth, and token length determine the information
capacity of tokenized representations. Existing methods commonly rely on
manually selected configurations
\citep{rajput2023recommender,liu2024discrete,
zhu2024cost,he2025learning}.
Adaptive quantization further allows different inputs to activate
different numbers of quantization levels
\citep{huijben2024residual,seo2024rate,chae2025variable}.
Related mechanisms have also been studied in flexible embeddings and
adaptive computation
\citep{kusupati2022matryoshka,graves2016act,
dehghani2019universal}. However, these approaches focus on input level
allocation and generally assume that the maximum available capacity has
already been specified.
\section{Preliminary}
\label{preliminary}

This section presents the overall research settings used in this paper, including the notation, the representation model pre-training pipeline which serves as the reference system for subsequent analyses, and the downstream evaluation protocol.

\subsection{Notation}

On a comprehensive internet platform such as Alipay, extensive nonsensitive user information and interactions are accessible. 
Let $\mathcal{U} = \{u_1, u_2, \ldots, u_N\}$ denotes the behavioral sequence dataset of $N$ users, and each user's behavior $u_n$ is organized according to the chronological order, such as daily, thus $u_n = \{ u^1_n, u^2_n, \ldots, u^D_n \}$ across $D$ days. The behavior $u^d_n$ contains multi-source data, for instance, PayBill $B$, Super Position Model (SPM) records $S$, and MiniProgram usage descriptions $M$, as $u^d_n = \{ B^d_n, S^d_n, M^d_n \}, n=1 \ldots N, d=1 \ldots D$. Information of each data source can be expressed in textual form, as $ B^d_n \in \mathcal{T}^{L_n^d(B)} $, $ S^d_n \in \mathcal{T}^{L_n^d(S)} $ and $ M^d_n \in \mathcal{T}^{L_n^d(M)} $ represent the raw text for PayBill, SPM and MiniProgram descriptions, respectively, where $ \mathcal{T} $ denotes the predefined text-token dictionary and $L_n^d(\cdot)$ represents the number of behavioral events recorded from each data source for user $n$ within the $d$-th time window ($L_n^d(\cdot)$ may vary across users and time windows due to heterogeneous activity levels).
Table \ref{tab:notation} summarizes the notations and corresponding description that used throughout this report.

\begin{table}[!t]
\centering
\caption{Notations, corresponding descriptions, and key statistics. Pretraining
    statistics are shown above the dashed line, and downstream
    evaluation settings are shown below.
}
\label{tab:notation}
\footnotesize
\setlength{\tabcolsep}{4pt}
\renewcommand{\arraystretch}{1.12}

\begin{tabularx}{\columnwidth}{
    @{}
    >{\raggedright\arraybackslash}p{0.15\columnwidth}
    >{\raggedright\arraybackslash}X
    >{\raggedright\arraybackslash}p{0.35\columnwidth}
    @{}
}
\toprule
\textbf{Notation}
&
\textbf{Description}
&
\textbf{Statistics}
\\
\midrule

$N$
&
Number of users
&
100M--2B
\\

$D$
&
Length of the user behavior sequence
&
30-270 days
\\

$P$
&
Model size of the pretrained user embedding model
&
0.05B--0.4B
\\

$u_n^d$
&
Behavior information of user $n$ within the $d$-th time window, including Paybill $B_n^d$, SPM $S_n^d$, and Miniprogram $M_n^d$
&
Daily events 0--10 per user; median $\sim$ 6
\\

$\mathbf{s}$
&
Behavior text tokens used for training
&
$\sim2200$ per user
\\

\hdashline

$\mathcal{D}_{test}$
&
Test dataset for downstream evaluation
&
Classification: 50 datasets, 0.5M users per set
\newline
Text retrieval: 22 datasets, 0.5M users per set
\newline
U2U retrieval: 22 datasets, 0.5M users per set
\\

\bottomrule
\end{tabularx}
\end{table}

\subsection{Model Pre-training Framework}

\begin{figure}[htbp]
    \centering
        \includegraphics[width=\linewidth]{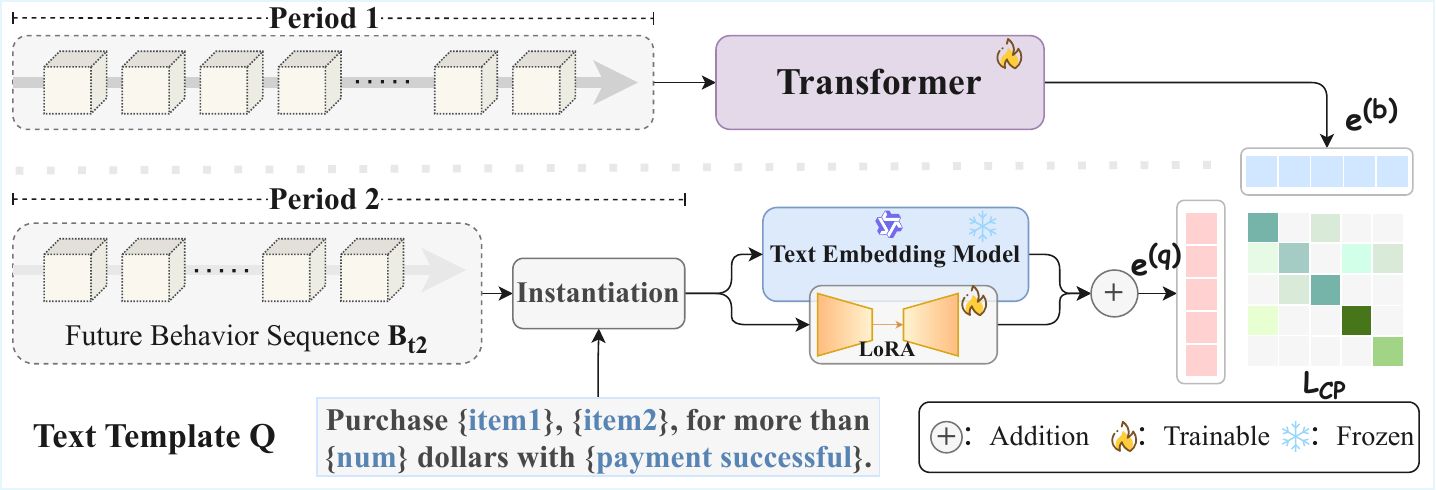}
        \caption{Pretraining framework for general-purpose user representation. Each user's behavioral sequence is split into a past and future segment, as the former part is encoded by a Transformer encoder and the latter is fed to a LoRA-tuned LLM-based model to generated embeddings, which are further aligned via a contrastive loss.}
    \label{fig:alignment}
\end{figure}

To establish a representation learning pipeline for user behavioral sequences, we utilize the self-supervised pre-training framework via behavior-text contrastive learning following \citep{FOUND}, as shown in Figure \ref{fig:alignment}.

Specifically, 
for each behavioral information item $u^d_n$, textual templates $\mathcal{Q}$ are designed to handle them to structured expression. It should be noticed that different templates are employed according to different data sources.
For example, the paybill information occurred by a user will be strung together into a complete description 
purchasing a certain product, belonging to a certain category, transaction amount and payment channel.
Templates take the form as:
\begin{tcolorbox}
\textbf{Template for purchase category}: The user purchased \{\textcolor{blue}{items}\} amounting more than \{\textcolor{blue}{num}\} dollars with \{\textcolor{blue}{status}\} payment.
\end{tcolorbox}

For pre-training, we first collect and curate corpus by splitting each behavioral sequence $u_n$ into two segments: historical part $ u_n^\textit{past} = (u^1_n, \dots, u^m_n) $ and future part $ u_n^\textit{future} = (u^{m+1}_n, \dots, u^D_n) $, where $m$ corresponds to the split time window boundary. 
Then a Transformer-based user encoder $f_\theta$ processes the past segment $u_n^\textit{past}$ to generate the embedding $\mathbf{e}_n^{b} \in \mathbb{R}^d$. 
Concurrently a textual description is instantiated from the template $\mathcal{Q}$ which takes the future segment $u_n^{\textit{future}}$ as input and sampled to generate the supervisory signal. 
This description is encoded by a LoRA\citep{lora}-tuned LLM-based embedding model $g_\phi$ (e.g., Qwen3-Embedding\citep{qwen3emb}) to produce an embedding $\mathbf{e}_n^{q} \in \mathbb{R}^d$. The base model is kept frozen and only low rank adapters with rank 16 is applied.
As for the pretraining objective, two embeddings are aligned via the contrastive loss Info-NCE\citep{cpc}, which can be denoted as:

\begin{equation}
\label{eq:contra_loss}
    \mathcal{L}_{\text{CP}} = -\frac{1}{B}\sum_{i=1}^{B} \log \frac{\exp(sim(\mathbf{e}_i^{b}, \mathbf{e}_i^{q}) / \tau)}{\sum_{j=1}^{B}\exp(\text{sim}(\mathbf{e}_i^{b}, \mathbf{e}_j^{q}) / \tau)}
\end{equation}
where $B$ means the training batch size, $sim$ represents the cosine similarity function and $\tau$ is a learnable temperature parameter. The denominator sums over all $B$ in batch pairs treating the description from the same user as positives and non matching description pairs as negatives.

\subsection{Downstream Evaluation Protocol}
\label{subsec:downstream_eval}

We evaluate the quality of learned user representations under different scaling settings using a fixed downstream evaluation protocol. 
The evaluation comprises three common tasks for user representation learning in real-world industrial application: (i) \textbf{Classification} (ii) \textbf{Text-based Retrieval} (iii) \textbf{U2U Retrieval}.
It can be noticed that evaluation datasets for each task span diverse scenarios, including user preference, risk control, recommendation, marketing, etc., and the reported values are the metric averages across all scenario-specific datasets to assess the performance and generalizability of the learned representation. 
Detailed dataset statistics are provided in the Table \ref{tab:notation}.

\begin{figure}[htbp]
    \centering
        \includegraphics[width=\linewidth]{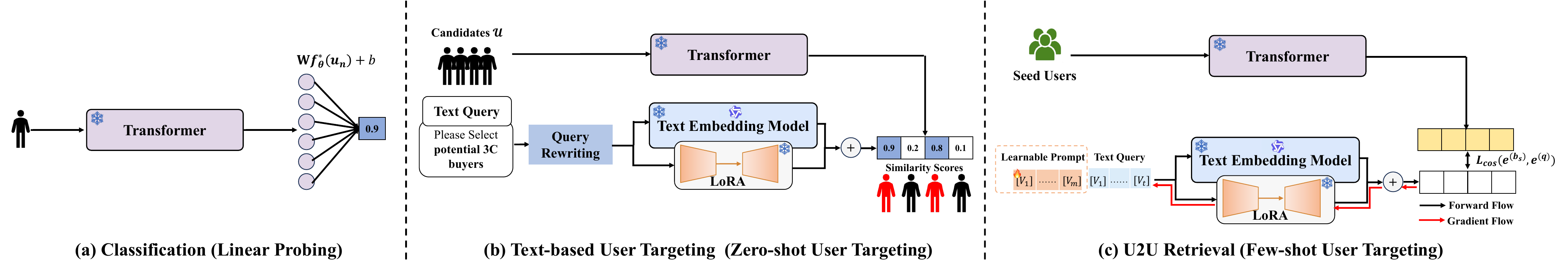}
        \caption{Three downstream tasks for user representation evaluation.}
    \label{fig:eval_task}
\end{figure}

\subsubsection{Classification}

Representation-based classification task is typically related to industrial scenarios such as advertising, marketing and risk control directly, and linear probing protocol is commonly adopted to evaluate the classification performance\citep{clip,cherti2023reproducible}. Specifically, we freeze the pretrained user encoder and train only a lightweight linear classifier to fit a given number of labeled samples
on top of the extracted embeddings. 

As for each pretrained user encoder $f^*_\theta$, we freeze and use it to extract user behavioral embeddings in all downstream datasets, which is further employed as:

\begin{equation}
\label{eq:contra_loss}
    y^{cls}_n = \mathbb{I}\! \left[ \mathbf{W} f^*_\theta(u_n) + b > \mathcal{T}^{cls} \right]
\end{equation}
, where $\mathbf{W} \in \mathbb{R}^{d},b \in \mathbb{R}$ is the linear network for binary classification which is trained and evaluated on the held-out training and test dataset, $\mathcal{T}^{cls}$ is the classification threshold commonly set at 0.5 .

\paragraph{Metrics.}
For the linear probing task, three complementary metrics are reported: \textbf{AUC}, \textbf{KS}, and \textbf{Accuracy}. 
\textbf{AUC (Area Under the ROC Curve)} measures the model’s overall ability to rank positive instances above negative ones across all classification thresholds. \textbf{KS (Kolmogorov–Smirnov)} value measures the maximum separation between the cumulative score distributions of positive and negative instances, reflecting the model’s strongest discriminative ability at an optimal threshold. \textbf{Accuracy} measures the proportion of correctly classified instances under a specified decision threshold, capturing overall prediction correctness. Together, these metrics evaluate threshold-independent ranking quality, class separability, and threshold-dependent classification performance, providing a comprehensive assessment of downstream classification effectiveness.

\subsubsection{Text-based Retrieval}
\label{subsubsec:text_retrieval}

User retrieval based on the text query constitutes a key component of the zero-shot user targeting task, which means judging whether each user belongs to the target audience corresponding to the given natural-language query $Q$.
As shown in Figure , 
by computing the cosine similarity between the text-query embedding and each candidate user embedding, users whose similarity scores exceed a predefined threshold are selected and targeted.

\begin{equation}
\label{eq:selected_audience}
    \begin{gathered}
    y_n^{tbr}
    =
    \mathbb{I}\!\left[
    sim\!\left(
    f^*_\theta(u_n),
    \mathcal{E}^*_\theta\!\left(\operatorname{Tok}[\mathcal{R}(q)]\right)
    \right)
    >
    \mathcal{T}^{tbr}
    \right],
    \\
    \mathcal{U}_q
    =
    \left\{
    u_n \in \mathcal{U}
    \;\middle|\;
    y_n^{tbr}=1
    \right\}
    \end{gathered}
\end{equation}
, where $sim$ denotes the cosine similarity function used during pretraining, $\mathcal{T}^{tbr}$ is the retrieval threshold (typically set to 0.5).

\paragraph{Metrics.}
\textbf{AUC}, \textbf{Precision} and \textbf{Recall} are employed to assess the text-based user retrieval task.
\textbf{AUC} measures the overall ranking quality between relevant and irrelevant users across all decision thresholds. \textbf{Precision} measures the proportion of selected users who are truly relevant to the targeting query, reflecting the reliability of the retrieval. \textbf{Recall} measures the proportion of all relevant users that are successfully retrieved, capturing the coverage of the targeting results. Together, these metrics assess ranking quality, targeting accuracy and coverage, providing a comprehensive evaluation of text-based user targeting performance.

\subsubsection{U2U Retrieval}

User-to-user retrieval based on a small set of seed users is a crucial step in building recommendation systems.
Therefore we also design an evaluation approach for the few-shot task as illustrated in Figure \ref{fig:eval_task} (c).
The few-shot retrieval task is accomplished via prompt-tuning.
Inspired by previous studies\cite{zhou2022learning}, the seed users $\mathcal{D}_{s}=\left\{x_i\right\}_{i=1}^{K}$ can be utilized as labels to learn contexts which improves the descriptive ability of the prompt, by adding learnable tokens $\mathbf{P}=[V_1,\ldots,V_m]$ to the input text.

\begin{equation}
\label{eq:few_shot_prompt_selection}
    \mathbf{P}^{*}
    =
    \underset{\mathbf{P}}{\arg\min}\;
    \mathcal{L}_{\mathrm{cos}}(
    f^*_\theta(u_n),
    f^*_\theta\!\left(\mathbf{P} \oplus \operatorname{Tok}(q)\right)
\end{equation}
, where $\mathcal{L}_{\mathrm{cos}}$ represents the cosine similarity loss.
Once trained, the tuned prompt can be used as the input to achieve user targeting via pipeline in Figure \ref{fig:eval_task} (b) following Eq \ref{eq:selected_audience}, as:

\begin{equation}
\label{eq:few_shot_prompt_selection}
    \begin{gathered}
    y_n^{u2ur}
    =
    \mathbb{I}\!\left[
    sim\!\left(
    f^*_\theta(u_n),
    \mathcal{E}^*_\theta\!\left(
    \mathbf{P} \oplus \operatorname{Tok}[q]
    \right)
    \right)
    >
    \mathcal{T}^{u2ur}
    \right],
    \\
    \mathcal{U}_q
    =
    \left\{
    u_n \in \mathcal{U}
    \;\middle|\;
    y_n^{u2ur}=1
    \right\}
    \end{gathered}
\end{equation}

\paragraph{Metrics.}
The same retrieval metrics \textbf{AUC}, \textbf{Precision} and \textbf{Recall} used in Section \ref{subsubsec:text_retrieval} are adopted.

\section{Scaling Law Analysis: Raw Data Scaling Wall}
\label{sec:scaling_analysis}

\begin{figure*}[htbp]
\centering
\includegraphics[width=\linewidth]{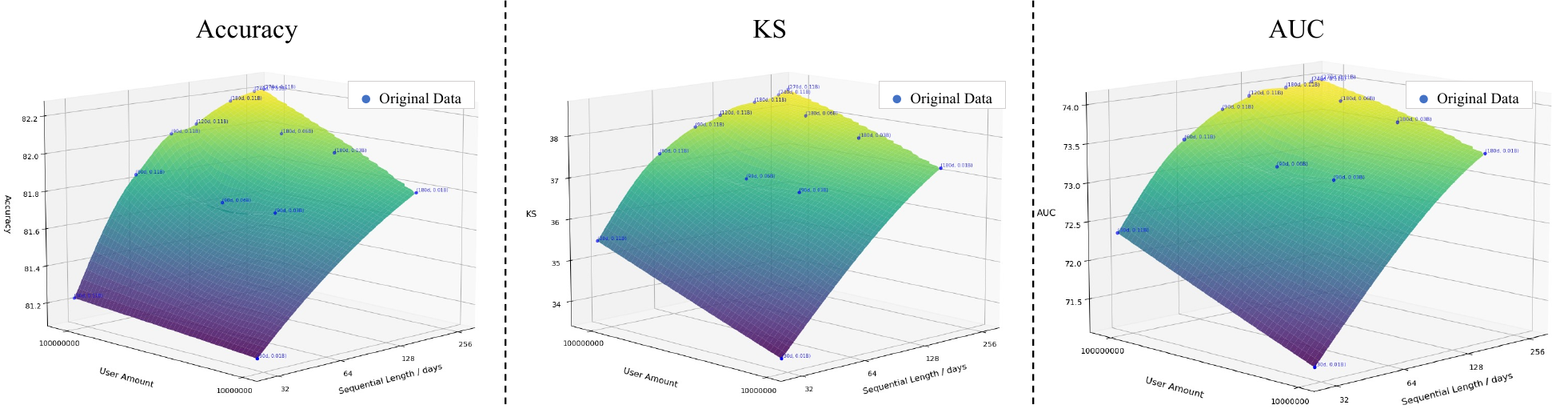}
\caption{Raw behavioral scaling saturates beyond approximately 0.03B users and 60 days. All three probing metrics exhibit a flattening trend around the saturation thresholds, indicating that additional raw data volume yields diminishing downstream returns.}
\label{fig:original_scaling_law}
\end{figure*}

Following the protocol established in Section~\ref{subsec:downstream_eval}, we empirically characterize the scaling behavior of user representations along three practical dimensions, namely temporal horizon $D$, user population $N$, and model capacity $M$. For each dimension, we evaluate downstream probing performance using the AUC, KS, and Accuracy metrics defined in Section~\ref{subsec:downstream_eval}. Figure~\ref{fig:original_scaling_law} visualizes the raw scaling surface across the $(N, D)$ grid, where a clear saturation pattern emerges. Across all three dimensions, we observe that simply increasing raw behavioral data or model size yields rapidly diminishing downstream returns. This demonstrates a raw behavioral scaling wall in industrial user representation learning, where data volume grows much faster than useful task relevant information.

\subsection{Saturation in User Population ($N$)}
\label{subsec:scaling_users}

\begin{bluebox}
\begin{center}
\emph{\textbf{Takeaway 1:} More users improve representation quality only before behavioral diversity is sufficiently covered. After that point, redundant behavioral patterns dominate the additional data.}
\end{center}
\end{bluebox}

We first examine whether increasing the number of users can continuously improve representation quality. To this end, we vary the user population $N$ from $0.01$B to $0.1$B while fixing the temporal horizon at $D=180$ days. The encoder size is fixed at approximately $0.1$B parameters.

As shown in Figure~\ref{fig:original_scaling_law}, increasing the user population initially improves downstream probing performance. When $N$ grows from $0.01$B to approximately $0.03$B, all three metrics show a clear upward trend. This indicates that adding more users in the low data regime introduces useful behavioral diversity and improves the separability of learned user representations.

However, the improvement quickly diminishes once $N$ exceeds approximately $0.03$B. Further increasing the user population to $0.1$B leads to only marginal performance gains. This flattening trend suggests that the major behavioral patterns have already been sufficiently covered. Beyond this point, additional users mainly contribute redundant signals rather than new task relevant information.

\subsection{Redundancy in Temporal Horizon ($D$)}
\label{subsec:scaling_horizon}

\begin{bluebox}
\begin{center}
\emph{\textbf{Takeaway 2:} Extending the behavioral window helps in the early stage, but long raw histories mainly introduce repetitive and low information events.}
\end{center}
\end{bluebox}

We next study whether longer behavioral histories provide more useful information for user representation learning. We vary the historical observation window $D$ from $30$ to $120$ days while fixing the user population at $N=0.1$B.

Figure~\ref{fig:original_scaling_law} shows that extending the temporal horizon also leads to diminishing returns. Increasing $D$ from $30$ to $60$ days provides a clear performance gain, suggesting that recent historical behaviors help the encoder capture more complete user preferences and consumption patterns.

Nevertheless, the benefit of longer histories becomes much weaker beyond $60$ days. Extending the observation window from $60$ to $120$ days brings only limited improvement in downstream metrics, despite doubling the amount of historical input. This suggests that long horizon PayBill sequences contain substantial repetition, including recurring payments and habitual consumption behaviors. As a result, additional history increases sequence length much faster than it increases downstream relevant information.
This saturation pattern is also reflected in the pretraining dynamics shown in Appendix Figure~\ref{fig:scaling_law_loss}. As the temporal horizon becomes longer, the contrastive alignment loss converges to increasingly similar final values, indicating that additional historical context does not provide proportionally more useful supervision.

\subsection{Quality Dissociation in Model Capacity ($P$)}
\label{subsec:scaling_model_parameters}

\begin{bluebox}
\begin{center}
\emph{\textbf{Takeaway 3:} Larger encoders achieve lower pretraining loss, but downstream accuracy saturates. This shows that information density, rather than model capacity alone, is the key bottleneck.}
\end{center}
\end{bluebox}

We further examine whether increasing model capacity can overcome the saturation caused by redundant behavioral data. We fix both the user population and the temporal horizon at $N=0.1$B and $D=180$ days, which already lies in the saturated data regime identified above. We then scale the Transformer-based user encoder from $0.05$B to $0.4$B parameters while keeping all other training configurations unchanged.

\begin{figure*}[htbp]
\centering
\includegraphics[width=\linewidth]{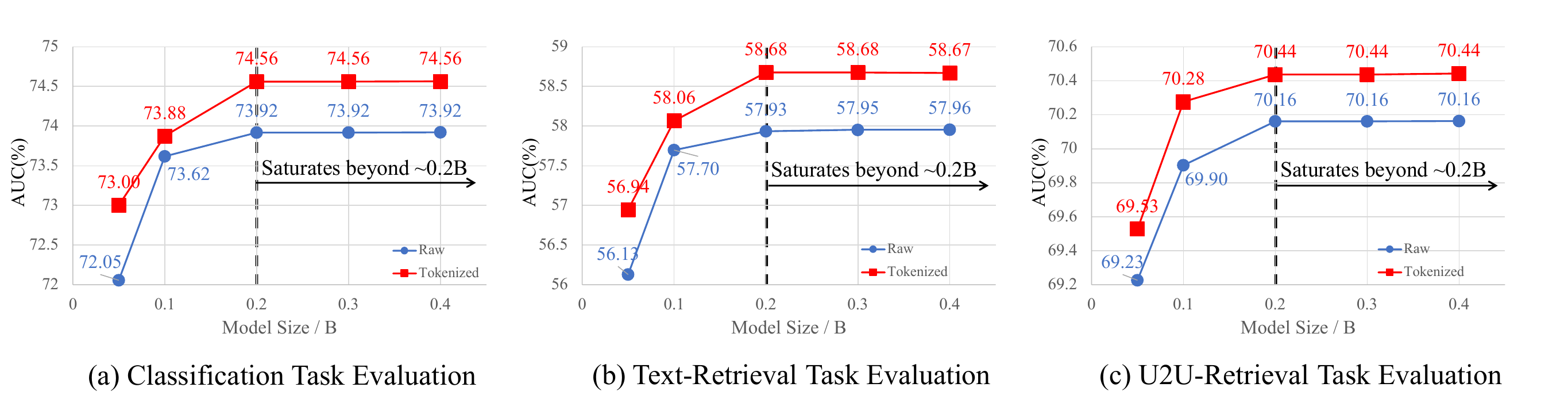}
    \caption{Scaling model parameters under fixed behavioral data exhibits strong diminishing returns. Larger models continue to achieve lower training loss, but downstream evaluation accuracy rapidly saturates beyond moderate model scales. This indicates that representation quality is constrained by information density rather than model capacity alone.}
\label{fig:capacity_trap}
\end{figure*}

Figure~\ref{fig:capacity_trap} shows a clear mismatch between optimization and downstream representation quality. Larger models consistently achieve lower training loss, confirming that increasing model capacity improves the ability to fit the pretraining objective. However, downstream evaluation shows a different trend. Increasing the model size from $0.05$B to $0.2$B improves test accuracy, but further scaling to $0.4$B produces almost no additional gain. Despite the larger parameter budget and higher computational cost, the learned user representations do not become more discriminative.

This result exposes a loss and quality dissociation. A lower pretraining loss does not necessarily translate into better downstream representation quality. In the saturated data regime, extra parameters mainly fit redundant behavioral details rather than learning more generalizable user features. Therefore, the limiting factor is not simply model expressivity, but the amount of downstream relevant signal contained in the raw behavioral input.

\subsection{Implications and Motivation for Behavioral Densing}
\label{subsec:motivation_for_densing}

\begin{bluebox}
\begin{center}
\emph{\textbf{Overall Takeaway:} The key to further scaling is no longer processing more raw behavioral events, but increasing the downstream relevant information carried by each effective input unit.}
\end{center}
\end{bluebox}

Taken together, the steepest performance improvements are confined to the region bounded by $N \le 0.03$B, $D \le 60$ days, and $M \le 0.2$B parameters. Beyond these thresholds, scaling user population, temporal horizon, or model capacity introduces substantial additional cost while yielding negligible downstream gains. These empirical results show that raw data volume and model size improve user representations only before redundancy becomes dominant.

These observations suggest that brute force scaling is \emph{not} a sustainable path for industrial user representation learning. Once raw behavioral data enters the saturation regime, further increasing data volume, compute, or model parameters leads to rapidly diminishing returns. The bottleneck is therefore not merely insufficient model capacity, but the low information density of raw behavioral sequences, where many events are repetitive, predictable, or only weakly relevant to downstream tasks.

This motivates a shift from volume oriented scaling to density oriented scaling. Rather than relying on more raw behavioral events, the goal is to make each effective input unit carry more downstream relevant signal under a fixed data and compute budget. The next section instantiates this idea through residual quantization. Instead of feeding long and repetitive raw sequences directly into the user encoder, we first transform them into compact discrete behavioral tokens, providing a practical mechanism for increasing behavioral information density and improving encoder utilization.

\section{Densing Gains from Tokenization}
\label{sec:densing}

We identify a raw behavioral scaling wall. Once redundancy dominates user histories increasing the user population extending temporal windows or enlarging model capacity yields diminishing downstream returns. This suggests that further improvement should come from increasing the information density of behavioral inputs rather than simply processing more raw events.

To operationalize this density oriented approach we employ residual quantization as a practical behavioral tokenizer. We adopt existing tokenization methods, e.g., RQ-VAE \citep{lee2022autoregressive,he2025learning}, as the quantization backbone 
for industrial user behavior modeling. Specifically we compress long multi source PayBill histories into compact discrete behavioral tokens and then train the same Transformer based user encoder on these tokenized sequences. This design allows us to test whether the gain comes from a denser input representation rather than from changes in the downstream encoder architecture.

\begin{figure*}[htbp]
\centering
\includegraphics[width=\linewidth]{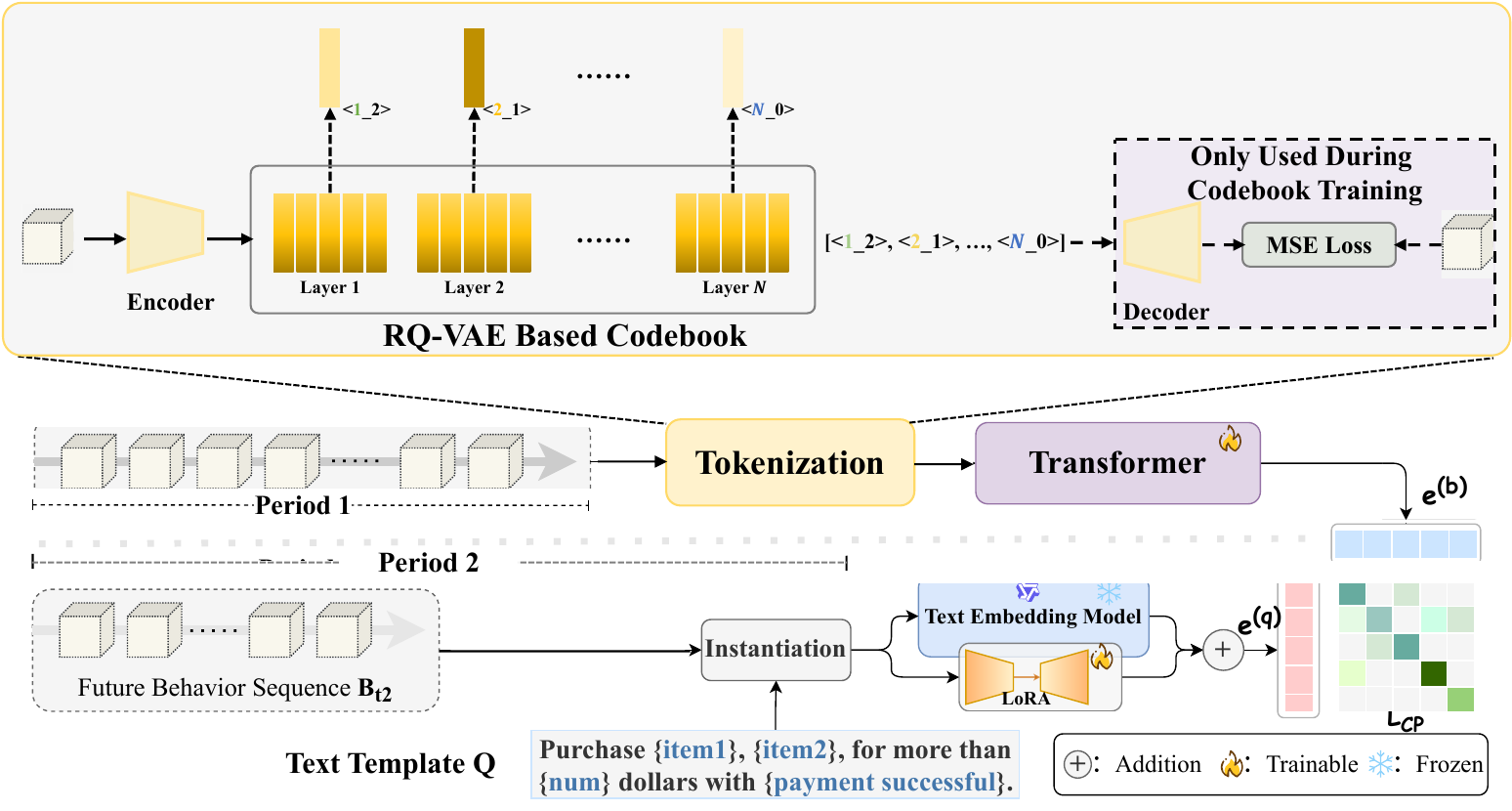}
\caption{Overview of the RQ-VAE based tokenization method used for behavioral densing.}
\label{fig:tokenized_encoder}
\end{figure*}

Figure~\ref{fig:tokenized_encoder} illustrates the overall pipeline. Raw multi source PayBill behaviors are first mapped into continuous behavioral embeddings. The RQ-VAE tokenizer then compresses these embeddings into multi level discrete tokens through residual quantization. The resulting token sequence is fed into the same user encoder and optimized with the same contrastive user text alignment objective as in the raw sequence setting.

\subsection{Behavioral Tokenization with RQ-VAE}
\label{subsec:rqvae_tokenization}

We now describe how residual quantization converts long and redundant behavioral histories into compact discrete tokens. To adapt RQ-VAE from image processing to industrial behavioral modeling we make three task specific modifications upfront. First instead of quantizing image latents the tokenizer operates on multi source PayBill behavior embeddings. Second we decouple the output token length $H$ from the raw temporal horizon $D$ so long histories can be represented by a fixed number of behavioral tokens. Third we use the tokenizer as a behavioral densing module for downstream user representation learning rather than only optimizing it for reconstruction quality.

Addressing the first modification given a user $u_n$ let $\mathbf{X}_n = (\mathbf{x}_{n,1}, \mathbf{x}_{n,2}, \ldots, \mathbf{x}_{n,L_n})$ denote the continuous embedding sequence derived from the raw behavioral history $B_n$ where each $\mathbf{x}_{n,\ell} \in \mathbb{R}^{d}$. Directly modeling $\mathbf{X}_n$ preserves all raw events but also retains substantial redundancy from repeated routines and weakly informative interactions.

To implement the second modification and increase representation density we apply a local aggregation step that maps the raw sequence into a compact latent representation $\mathbf{Z}_n = (\mathbf{z}_{n,1}, \mathbf{z}_{n,2}, \ldots, \mathbf{z}_{n,H})$ where $H \ll L_n$ is the fixed token budget of the behavioral tokenizer. This step decouples the effective input length from the raw temporal horizon and encourages the tokenizer to retain the most informative behavioral semantics under a compact budget.

For each latent vector $\mathbf{z}_{n,h}$ RQ-VAE applies multi stage residual quantization. At the $m$-th stage a learned codebook $\mathcal{C}^{(m)}=\{\mathbf{c}^{(m)}_1,\ldots,\mathbf{c}^{(m)}_{K_m}\}$ quantizes the current residual by selecting its nearest codeword. The process can be written compactly as
\begin{equation}
\label{eq:rq_tokenization}
\begin{aligned}
\mathbf{r}_{n,h}^{(1)} &= \mathbf{z}_{n,h}, \\
k_{n,h}^{(m)} &=
\arg\min_{k \in \{1,\ldots,K_m\}}
\left\|
\mathbf{r}_{n,h}^{(m)} - \mathbf{c}^{(m)}_k
\right\|_2^2, \\
\mathbf{r}_{n,h}^{(m+1)}
&=
\mathbf{r}_{n,h}^{(m)}
-
\mathbf{c}^{(m)}_{k_{n,h}^{(m)}}, \\
\hat{\mathbf{z}}_{n,h}
&=
\sum_{m=1}^{M}
\mathbf{c}^{(m)}_{k_{n,h}^{(m)}}, \\
\mathbf{t}_{n,h}
&=
\big(k_{n,h}^{(1)}, k_{n,h}^{(2)}, \ldots, k_{n,h}^{(M)}\big).
\end{aligned}
\end{equation}
The full user history is therefore represented as a compact token sequence $\mathbf{T}_n = (\mathbf{t}_{n,1}, \mathbf{t}_{n,2}, \ldots, \mathbf{t}_{n,H})$.

This residual formulation provides a natural coarse to fine structure for behavioral modeling. Early codebooks capture dominant macro level semantics such as stable consumption categories routine transportation or frequent payment scenarios. Later codebooks refine residual variations such as merchant level preferences or subtle changes in spending behavior which may still be predictive for downstream targeting. Compared with a single codebook residual quantization can preserve finer behavioral distinctions under the same compact token budget.

The tokenizer is trained with a reconstruction objective and a commitment loss
\begin{equation}
\label{eq:rq_loss}
\mathcal{L}_{\mathrm{RQ}}
=
\mathcal{L}_{\mathrm{rec}}(\mathbf{Z}_n, \hat{\mathbf{Z}}_n)
+
\sum_{m=1}^{M}
\left\|
\mathrm{sg}[\mathbf{r}^{(m)}] - \hat{\mathbf{r}}^{(m)}
\right\|_2^2
+
\beta
\sum_{m=1}^{M}
\left\|
\mathbf{r}^{(m)} - \mathrm{sg}[\hat{\mathbf{r}}^{(m)}]
\right\|_2^2 
\end{equation}
where $\mathcal{L}_{\mathrm{rec}}$ denotes the reconstruction loss $\mathrm{sg}[\cdot]$ is the stop gradient operator and $\beta$ controls the commitment penalty.

Reflecting the third modification after tokenizer pretraining we freeze the RQ-VAE and train the user encoder on $\mathbf{T}_n$ using the same alignment objective as the raw sequence baseline. This separation allows us to attribute downstream gains strictly to behavioral densing rather than additional encoder capacity.

\subsection{Experimental Evaluation of Discrete Tokenization}
\label{subsec:tokenization_evaluation}

\begin{figure*}[htbp]
\centering
\includegraphics[width=\linewidth]{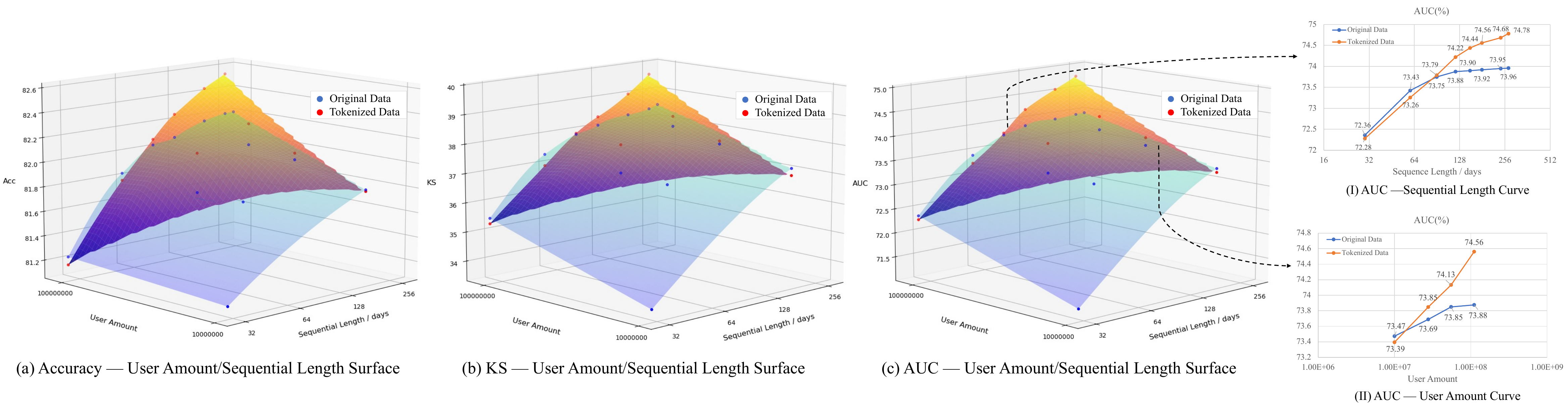}
\caption{Tokenized representations outperform raw sequences at matched data scales with the gap widening where raw modeling saturates. Subplot (a) shows the full two dimensional scaling surface. Subplots (b) and (c) show cross sections at fixed user population and fixed temporal horizon.}
\label{fig:tokenized_scaling_law}
\end{figure*}

To evaluate whether residual quantization alleviates the raw behavioral scaling wall we repeat the scaling protocol from Section~\ref{sec:scaling_analysis}. We compare raw sequence modeling with tokenized sequence modeling under matched data and compute budgets. In the raw setting the user encoder is trained directly on PayBill sequences. In the tokenized setting raw histories are first converted into RQ-VAE tokens and the same user encoder is trained on the resulting discrete token sequences. All models use the same contrastive user text alignment objective and are evaluated with the same downstream probing tasks and metrics.

Figure~\ref{fig:tokenized_scaling_law} compares the two settings across different scales. Along the temporal dimension tokenized representations become increasingly advantageous as the observation window grows longer. At shorter horizons raw sequences remain competitive because behavioral redundancy is still limited. Around $D \approx 64$ days the tokenized curve begins to outpace the raw baseline and the advantage becomes clearer as the horizon further increases. This indicates that residual quantization better distills useful behavioral signal from long and repetitive histories.

A similar pattern appears along the user population dimension. At smaller user scales raw and tokenized models perform similarly since adding users still introduces useful behavioral diversity. As the user population grows and raw scaling approaches saturation the tokenized representation shows a clearer advantage. In our experiments this crossover appears around $N \approx 1.2 \times 10^{7}$ after which the tokenized representation consistently outperforms the raw baseline. This suggests that behavioral densing is most helpful when the marginal utility of additional raw data becomes low.

To quantify this effect let $\mathcal{P}_{\mathrm{raw}}(s)$ and $\mathcal{P}_{\mathrm{tok}}(s)$ denote downstream probing performance under raw and tokenized modeling where $s$ can represent temporal horizon $D$ or user population $N$. We define the tokenization gain as
\begin{equation}
\label{eq:density_gain}
\Delta(s)
=
\mathcal{P}_{\mathrm{tok}}(s) - \mathcal{P}_{\mathrm{raw}}(s).
\end{equation}
Empirically $\Delta(s)$ is not uniformly positive at all scales. Instead it becomes positive and grows larger when raw modeling enters the diminishing return regime. This pattern confirms that RQ-VAE tokenization does not merely shorten the input. It increases the effective density of behavioral information by suppressing repetitive details and preserving task relevant distinctions.

\begin{figure*}[!ht]
    \centering
    \includegraphics[width=\linewidth]{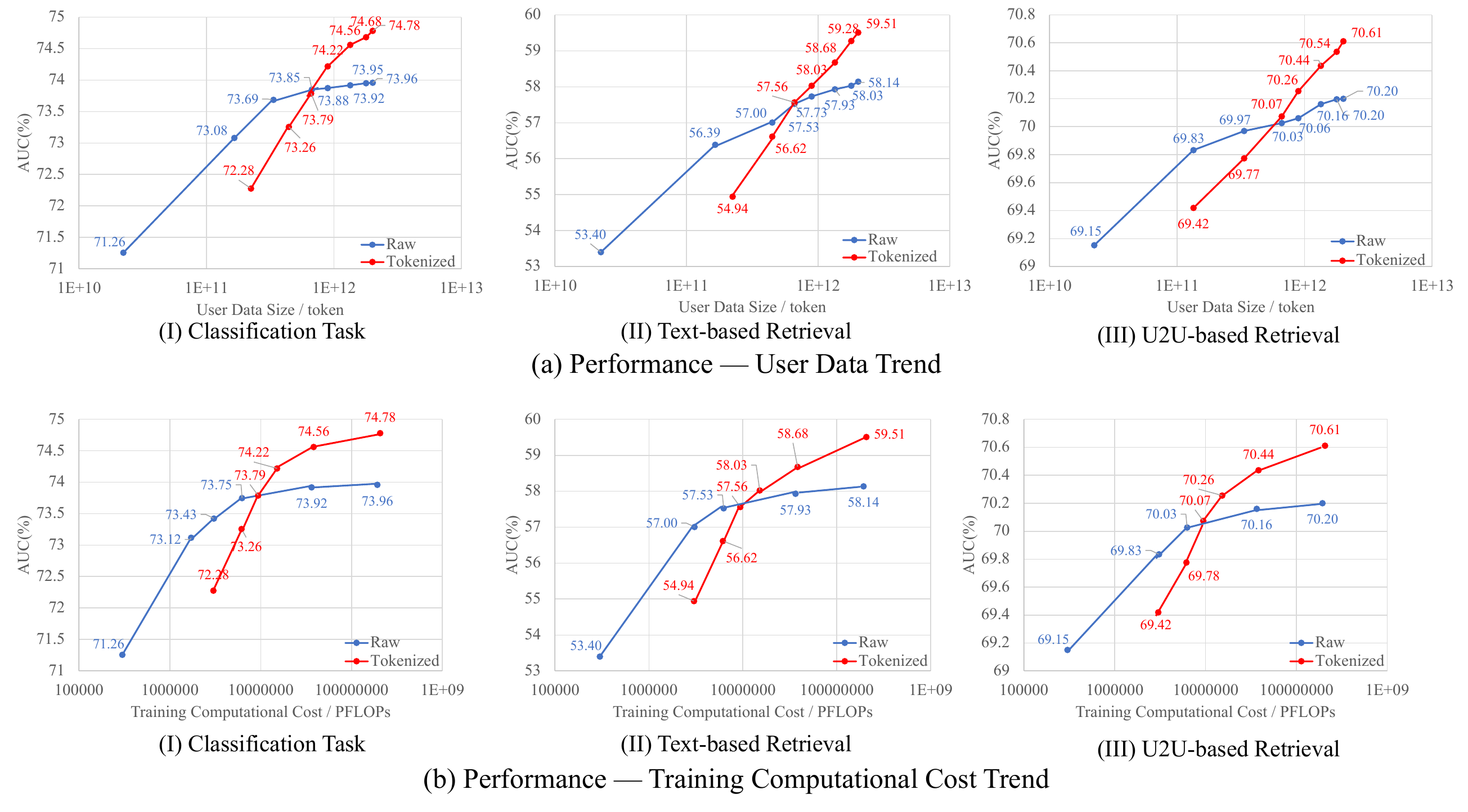}
    \caption{
    Downstream task performance with data size \& training computes scaling.}
    \label{fig:encoder_scaling}
\end{figure*}

\subsection{From Raw Scaling to Density Scaling}
\label{subsec:density_scaling_discussion}

The empirical comparison supports the central motivation of behavioral densing. Once raw behavioral data reaches saturation representation quality is no longer primarily limited by the volume of users the length of the historical window or the size of the encoder. Instead the fundamental limiting factor is the amount of downstream relevant signal contained in each effective input unit. 
Here RQ-VAE tokenization acts as a practical density operator. It shifts the learning problem from modeling long redundant raw sequences to modeling compact and semantically concentrated tokens. By suppressing repetitive routines and preserving task relevant distinctions the tokenizer enables moderate capacity encoders to better utilize long behavioral histories without a proportional increase in computational burden.

However these results also reveal a limitation of fixed length tokenization. Behavioral information is naturally heterogeneous. Some history segments contain rich and diverse signals while others consist mostly of predictable routines. A uniform token budget across all segments is therefore suboptimal. This observation directly motivates the next section where we allocate token capacity adaptively according to behavioral complexity rather than applying a fixed sequence length.

\section{The Behavioral Densing Law}
\label{sec:densing_law}

We reveal that raw behavioral scaling eventually saturates and that tokenization improves performance precisely in this saturated regime. This benefit reflects a fundamental density principle where an effective representation must maximize downstream relevant information while minimizing the cost of modeling redundant histories. 

We formalize this principle as the Behavioral Densing Law. Rather than treating densing as a heuristic configuration table we formulate it as a performance and cost Pareto optimization problem. The law characterizes how the Pareto optimal representation capacity evolves with behavioral scale. RQ-VAE tokenization serves as one concrete solver for estimating this trajectory under large scale behavior data.

\begin{figure*}[htbp]
\centering
    \includegraphics[width=\linewidth]{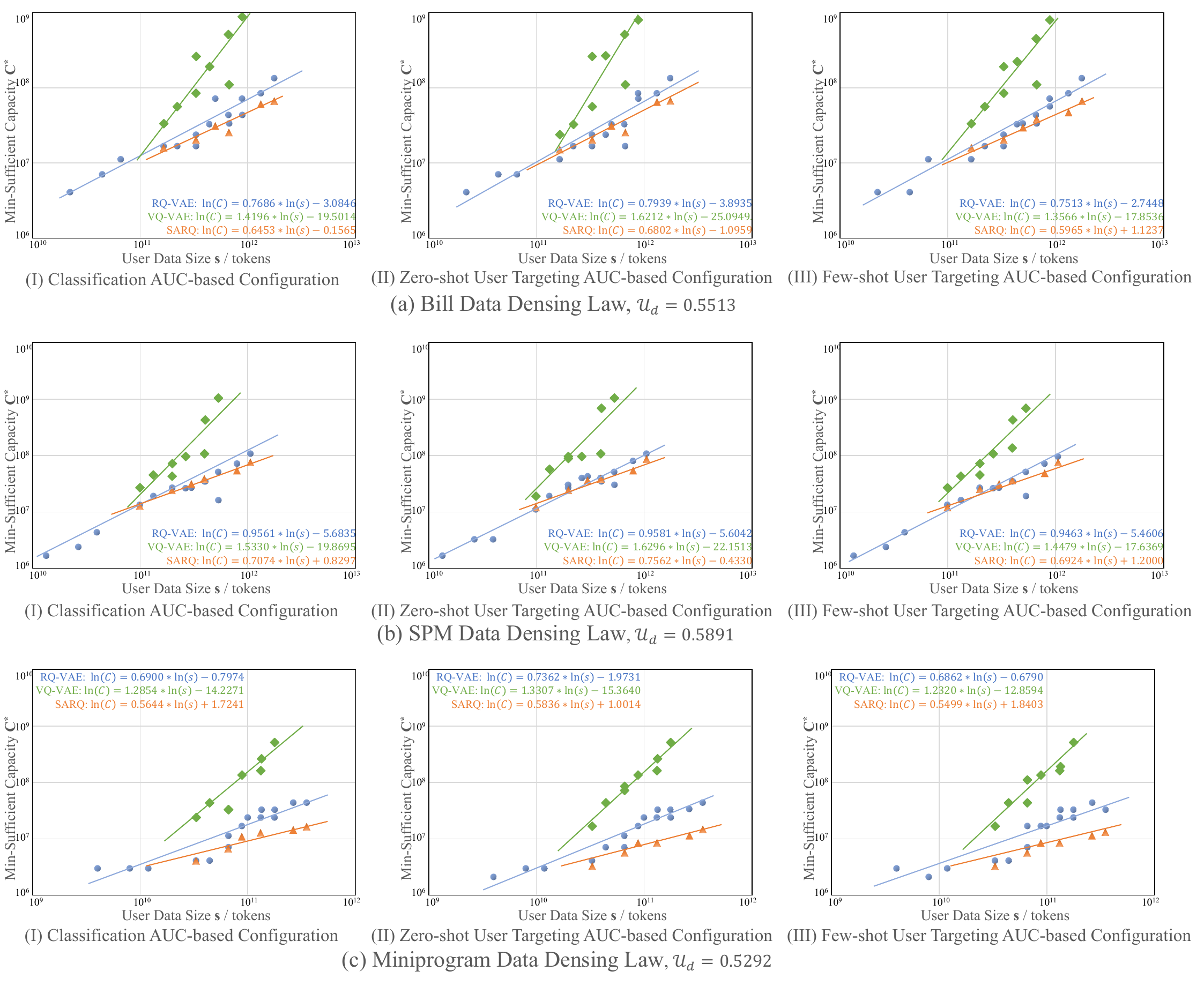}
    \caption{Behavioral Densing Law across different data sources and tokenization methods, derived from three  tasks.}
    \label{fig:densing_law_diffsrc}
\end{figure*}

\subsection{Behavioral Density and Pareto Optimal Densing}
\label{subsec:pareto_densing}

To explicitly quantify this principle we define behavioral density. Let $\mathbf{s} = (s_1, s_2, \ldots, s_r)$ denote a generalized behavioral scale state where each component captures one source of behavioral complexity such as user population temporal horizon modality count or scenario diversity. 

Let $C(\phi)$ be the actual cost of a densified representation $\phi$ and let $\widehat{C}_{\mathrm{raw}}(\mathbf{s})$ denote the effective raw budget which is the cost required for raw sequence modeling to achieve the same downstream utility. The optimal behavioral density at scale state $\mathbf{s}$ is defined as
\begin{equation}
\rho^*(\mathbf{s}) = \frac{\widehat{C}_{\mathrm{raw}}(\mathbf{s})}{C^*(\mathbf{s})}.
\end{equation}
The goal of behavioral densing is to maximize this density ratio navigating the representation toward a better performance and cost frontier.

At a given scale state $\mathbf{s}$ each representation transformation $\phi$ yields a downstream utility $U(\phi, \mathbf{s})$ and incurs a representation cost $C(\phi)$. Introducing a deployment tradeoff parameter $\lambda$ the Pareto optimal representation is formulated as
\begin{equation}
\phi^*(\lambda, \mathbf{s})
=
\arg\max_{\phi \in \Phi}
\left[
U(\phi, \mathbf{s})
-
\lambda C(\phi)
\right].
\label{eq:pareto_objective}
\end{equation}

The parameter $\lambda$ controls the cost sensitivity of the system. A larger $\lambda$ favors cheaper representations while a smaller $\lambda$ allows higher capacity when it brings sufficient downstream gain. Consequently densing does not aim to maximize the compression ratio or the tokenizer size. It specifically targets the representation with the best utility and cost tradeoff at a given scale. We define the Behavioral Densing Law as the solution map induced by this Pareto objective.

\subsection{Minimal Sufficient Tokenization Capacity}
\label{subsec:minimal_sufficient_capacity}

We instantiate this theoretical solution using different tokenization methods. Let $q_\kappa$ denote an RQ-VAE tokenizer configured by $\kappa = (K, M, d, H)$ where $K$ is the codebook size $M$ is the number of residual quantization stages $d$ is the code embedding dimension and $H$ is the output token length. 

For RQ-VAE tokenization we instantiate the general representation cost $C(\phi)$ with the discrete tokenizer capacity $C_{\mathrm{tok}}(\kappa)$. We quantify this capacity as
\begin{equation}
C_{\mathrm{tok}}(\kappa)
=
H M \log K
+
\eta H M d,
\label{eq:tokenizer_capacity}
\end{equation}
where the first term measures discrete code capacity and the second term accounts for continuous code embedding cost. 

For a given scale $(N, D)$ the theoretical Pareto objective is to find the optimal configuration $\kappa^*$. However downstream performance is measured with stochastic training noise making exact continuous optimization over $\lambda$ unstable. We therefore approximate the Lagrangian Pareto objective using a robust constrained form. We first compute the near optimal utility
\begin{equation}
U^*(N, D)
=
\max_{\kappa \in \mathcal{K}}
U(q_\kappa, N, D),
\end{equation}
and then rigorously select the smallest configuration that reaches this utility bound
\begin{equation}
\kappa^*(N, D)
=
\arg\min_{\kappa \in \mathcal{K}}
C_{\mathrm{tok}}(\kappa)
\quad
\mathrm{subject \ to}
\quad
U(q_\kappa, N, D)
\ge
U^*(N, D) - \epsilon.
\label{eq:minimal_capacity}
\end{equation}

This constrained form selects a Pareto efficient point that is robust to stochastic training noise and discrete tokenizer configurations. It safely identifies the minimal sufficient capacity avoiding larger configurations whose additional capacity provides negligible downstream gain.

\subsection{The Induced Scale Law}
\label{subsec:induced_scale_law}

To obtain a compact empirical form of this solution map we study the trajectory of Pareto optimal solutions across varying scale states. In the redundancy dominated regime the utility frontier of tokenized representations exhibits diminishing returns with respect to representation capacity $C$. A local approximation of this frontier is
\begin{equation}
U(C, \mathbf{s})
=
U_\infty(\mathbf{s})
-
a(\mathbf{s})C^{-b},
\label{eq:saturating_frontier}
\end{equation}
where $U_\infty(\mathbf{s})$ is the maximum attainable utility at scale $\mathbf{s}$ while $a(\mathbf{s})$ measures the remaining utility gap at limited capacity and $b>0$ controls the curvature of the frontier.

Substituting Equation~\ref{eq:saturating_frontier} into the continuous Pareto objective yields
\begin{equation}
C^*(\lambda, \mathbf{s})
=
\arg\max_C
\left[
U(C, \mathbf{s})
-
\lambda C
\right].
\end{equation}
The first order condition provides
\begin{equation}
a(\mathbf{s}) b \left(C^*(\lambda, \mathbf{s})\right)^{-b-1}
=
\lambda,
\end{equation}
which solves to
\begin{equation}
C^*(\lambda, \mathbf{s})
=
\left(
\frac{a(\mathbf{s})b}{\lambda}
\right)^{\frac{1}{b+1}}.
\label{eq:solved_capacity}
\end{equation}

Assuming the scale dependent gap term grows according to a general power law $a(\mathbf{s}) = a_0 \prod_{i=1}^{r} (s_i/s_{i,0})^{\gamma_i}$ we can rewrite Equation~\ref{eq:solved_capacity} as
\begin{equation}
\ln C^*(\lambda, \mathbf{s})
=
\beta
+
\sum_{i=1}^{r} \alpha_i \ln \frac{s_i}{s_{i,0}}
-
\alpha_\lambda \ln \lambda,
\label{eq:capacity_law_from_pareto}
\end{equation}
where $\alpha_i = \gamma_i / (b+1)$. In practical deployments the tradeoff $\lambda$ typically remains fixed. Equation~\ref{eq:capacity_law_from_pareto} thus reduces to the general operational form of the Behavioral Densing Law
\begin{equation}
\boxed{
\ln C^*(\mathbf{s})
=
\beta
+
\sum_{i=1}^{r} \alpha_i \ln \frac{s_i}{s_{i,0}}
}
\label{eq:general_densing_law}
\end{equation}

This derivation provides the scale dependent form of Pareto optimal capacity under a diminishing return utility frontier. It dictates that the required capacity of the optimal densified representation grows as a power law function of behavioral complexity.

In our scaling experiments the behavioral scale state is specifically instantiated as $\mathbf{s} = (N, D)$ where $N$ represents the user population and $D$ represents the temporal horizon. Equation~\ref{eq:general_densing_law} therefore reduces to the task specific empirical form
\begin{equation}
\ln C^*(N, D)
=
\beta
+
\alpha_N \ln \frac{N}{N_0}
+
\alpha_D \ln \frac{D}{D_0}
\label{eq:capacity_law_fixed_lambda}
\end{equation}
This confirms that $N$ and $D$ are not strict prerequisites for the law but rather two controllable dimensions used in this work to parameterize behavioral scale.

We refer to
Equation~\ref{eq:general_densing_law} as the \emph{Behavioral Densing Law}.
In the scalar setting, it becomes

\begin{equation}
\boxed{
    \ln C^*(s)=\beta+\alpha\ln(s/s_0)
}
\label{eq:general_densing_law}
\end{equation}
.

For further analysis, 
the coefficients $\alpha_i$ 
should summarize the capacity trajectory under a particular behavioral data
distribution and code-space expression form,
which can be represented as
\begin{equation}
\alpha_i
=
f_i
\left(
\mathcal{U}_d,
\mathcal{E}_{\phi}
\right),
\label{eq:densing_coefficient_factors}
\end{equation}
where $\mathcal{U}_d$ captures data properties such as behavioral
diversity and redundancy, 
while
$\mathcal{E}_{\phi}$ captures how tokenization method $\phi$ expresses the
available code space, including codebook organization, residual depth, and
capacity allocation.
It's analyzed that the intra-datasource diversity should be positively correlated with the tokenization capacity, as richer information should be mapped to a larger representation space to avoid conflicts.
In this study, for the quantitative description of the property, we utilize the \textit{Mean k-NN Cosine Distance} of LLM-based text embedding (e.g., Qwen3-Embedding) as the measurement, which can be denoted as:
\begin{equation}
    \mathcal{U}_d = \frac{1}{Mk}\sum_{i=1}^{M}
    \sum_{j \in \operatorname{kNN}(i)} d(i,j)
\label{eq:unique_def}
\end{equation}

\subsection{Empirical Validation of the Solution Trajectory}



    

Figure \ref{fig:densing_law_diffsrc} reports the empirical minimal sufficient tokenization configurations under varying measured data scales. The measured configurations display a clear structural pattern. As either the user population or the temporal horizon increases the minimal sufficient tokenization capacity also increases monotonically. 

By fitting Equation~\ref{eq:capacity_law_fixed_lambda} over these empirical configurations we estimate the scaling coefficients 
in Figure \ref{fig:densing_law_diffsrc} for different data sources.
This quantitative result confirms our theoretical formulation. In the redundancy dominated regime the Pareto optimal capacity of behavioral tokenization grows strictly as a power law function of user population and temporal horizon. 

Crucially this does not imply that larger tokenizers are unconditionally better. Instead it dictates that larger behavioral scales require a larger minimal sufficient capacity. Exceeding this optimal capacity yields negligible performance benefits while severely degrading cost efficiency. The measured configurations and the resulting fitted coefficients thus represent the empirical solution trajectory of the Behavioral Densing Law estimated under large scale behavioral data.

We also estimate coefficients from empirical minimal-capacity trajectories, and obtain the approximate relation between  $\alpha_i$ and $\mathcal{U}_d$ as $\alpha_i \propto \mathcal{U}_d^{2}$,
as we conducted a series of experiments to figure the minimal sufficient configurations tailored to each input user data size and utilize the Pareto-Optimal method to fit the law (similar to \citep{cherti2023reproducible}).

To ensure generalizability and reliability, the validation experiments are performed on three tokenization methods: RQ-VAE, VQ-VAE and SARQ, and three different user data sources: Paybill, SPM and Miniprogram.
We also evaluated all the three tasks in Sec \ref{subsec:downstream_eval}.
For each tokenization method's capacity configuration, size of the representation space for all available SID is calculated, as for RQ-VAE it's obtained by $C_{RQ-VAE}=K^M$, $C_{VQ-VAE}$ is determined by the searching space specified during training for VQ-VAE, and $C_{SARQ}$ is the activated SID capacity space for SARQ.
For the optimal configuration $C^*$, searching approach in Section \ref{subsec:pareto_densing} is utilized.

The experimental validation results are presented in Figure \ref{fig:densing_law_diffsrc}. According to the results we summarize the following patterns:

\paragraph{Approximately Linear Relationship on the Logarithmic Input Scale.}
Across all input data sources and evaluation tasks, the optimal capacity and input user data size exhibit an approximately linear relationship on a logarithmic scale ($ln(C^*)$ and $ln(s)$), as indicated by the distribution of the data points.

\paragraph{Identical Trend for Different Downstream Tasks.}
It's observed that the trends of the minimal sufficient configuration and input data size share the similar pattern under all tasks.
At the given tokenization method and data source (the same color within Figure \ref{fig:densing_law_diffsrc} (a), (b) or (c)), the slopes between $ln(C^*)$ and $ln(s)$ obtained from the Pareto frontier in (I), (II) and (III) can be considered identical within the numerical estimation error. 

\paragraph{Slope Affected by the Tokenization Method.}
For different tokenization approaches, the fitted distribution varies, as illustrated in different color lines in each subfigure.
It's also observed that VQ-VAE retains higher slope value compared to RQ-VAE, and SARQ reaches the lowest value. 
Our analysis suggests that this is related to each method's representation space redundancy $\mathcal{E}_{\phi}$. VQ-VAE suffers from redundant storage, as similar representations require multiple complete SID and wastes shared structures. In contrast, RQ-VAE enables compositional reuse of SID
thereby reducing $\mathcal{E}_{\phi}$ in the codebook. However, among the theoretically possible \(K^M\) combinations in RQ-VAE, many are invalid or have an extremely low probability of being selected,
which can be alleviated to some extent in SARQ.

\paragraph{Slope Proportional to Intra-Source Uniqueness.}
By comparing the slopes of the same colored line from Figure \ref{fig:densing_law_diffsrc}'s column subfigure (e.g.,(I) in (a), (b) and (c)),
we also observe that within the same tokenization method, the slopes probably retain the approximately proportional relation corresponding to the squared value of intra-datasource uniqueness $\mathcal{U}_{d}$, as the rate between three data sources is around $1:1.15:0.92$ while the $\mathcal{U}_{d}$ rate is around $1:1.07:0.96$ (within $10^{-2}$ numeric error).
This discovery provides guidance for tokenization-based user representation learning, which means when using the common tokenization method (with fixed $\mathcal{E}_{\phi}$), the optimal codebook capacity configuration can be obtained using Eq \ref{eq:general_densing_law} and Eq \ref{eq:densing_coefficient_factors} with the calculation of $\mathcal{U}_{d}$ according to Eq \ref{eq:unique_def}.

\section{New Method: Adaptive-Length-Gating-Based RQ-VAE Quantization}
\label{sec:new_token_method}

Following the Behavioral Densing Law in Section~\ref{sec:densing_law}, we further move from scale-level capacity selection to instance-level capacity allocation. The law suggests that the optimal densified representation should use the minimal sufficient capacity required by the input token cost. However, fixed-length tokenization methods, including VQ-VAE and RQ-VAE based tokenizers, assign the same number of code levels to all behavioral periods. This uniform allocation ignores the heterogeneity of user behaviors. Some periods contain diverse and ambiguous behavioral signals and require deeper residual codes, while others mostly consist of routine or repetitive behaviors and can be represented with shorter codes.

This motivates a variable-length tokenization strategy. The key principle is simple: a new residual code should be used only when its expected marginal utility is larger than its marginal representation cost. Therefore, instead of choosing a single global code length for all behavioral periods, we adaptively decide whether each behavioral representation should continue to the next quantization level. This provides an instance-level realization of the Behavioral Densing Law.

\subsection{Method}
\label{sec:new_token_method}

\begin{figure*}[!ht]
  \centering
  \includegraphics[width=\linewidth]{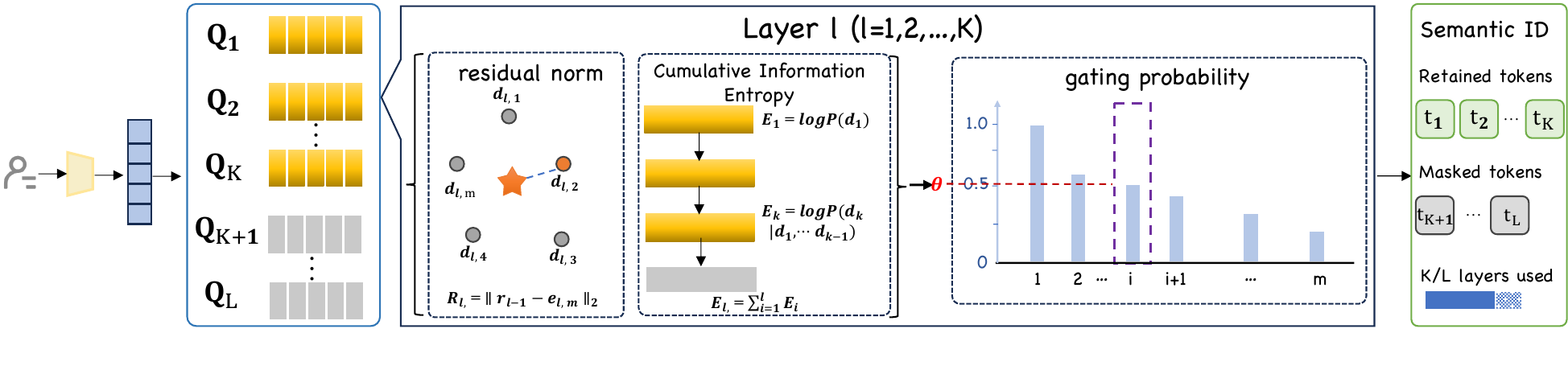}
  \caption{Adaptive variable-length tokenization method allocates more codes to high-information behavioral periods and fewer codes to routine or redundant periods, achieving comparable or better accuracy with fewer tokens on average than fixed-length baselines.}
  \label{fig:new_token_method}
\end{figure*}

To implement this instance-level capacity allocation principle, as shown in Figure~\ref{fig:new_token_method}, we propose an \emph{Adaptive Length Gated Network} (ALGN) to generate variable-length semantic IDs for user behavioral data. ALGN is built on top of the residual quantization architecture. At each quantization level, it decides whether the current behavioral representation still needs an additional residual code.

The design of ALGN follows a marginal utility view. For a behavioral representation at level $l$, continuing to the next residual code is beneficial only if the remaining information is still large enough to justify the extra code cost. Since the true downstream marginal utility is not directly observable during tokenization, we approximate it using two signals.

First, the residual norm measures the amount of information that remains unexplained by the current code path. Let $r_{l-1}$ denote the residual before level $l$, and let $e_{l,m}$ denote the selected code embedding at this level. The residual norm is defined as
\begin{equation}
    R_l
    =
    \left\|
    r_{l-1} - e_{l,m}
    \right\|_2 .
\label{eq:res_norm}
\end{equation}
A larger $R_l$ indicates that the current codes are still insufficient to reconstruct the behavioral representation, so an additional residual code may bring meaningful utility. A smaller $R_l$ suggests that the remaining information has already been mostly explained, and continuing the quantization process may lead to diminishing returns.

Second, the code uncertainty measures how ambiguous the current semantic assignment is. Let $p(m_l \mid m_{<l})$ denote the conditional probability of selecting code $m_l$ given previous code selections. We define the cumulative uncertainty as
\begin{equation}
    E_l
    =
    -
    \sum_{k=1}^{l}
    \log p(m_k \mid m_{<k}) .
\label{eq:entropy}
\end{equation}
A larger $E_l$ means that the code path is more uncertain, which usually corresponds to more complex, ambiguous, or information-rich behavioral patterns. A smaller $E_l$ indicates that the tokenizer is confident about the current semantic path, so additional code levels are less necessary.

Based on these two proxies, ALGN estimates the continuation probability at level $l$ as
\begin{equation}
    g_l
    =
    \sigma
    \left(
    \mathrm{sp}(w_R) R_l
    +
    \mathrm{sp}(w_E) E_l
    -
    \mathrm{sp}(w_C) c_l
    +
    b
    \right),
\label{eq:continue_rate}
\end{equation}
where $w_R$, $w_E$, $w_C$, and $b$ are learnable parameters, $\sigma(\cdot)$ is the sigmoid function, and $\mathrm{sp}(\cdot)$ is the Softplus function used to keep the weights positive. Here $c_l$ denotes the marginal cost of activating level $l$, such as the cumulative code length or level index. This form encodes the intended monotonic behavior: the continuation probability increases when the residual information or uncertainty is high, and decreases when the additional code cost becomes large.

The activated semantic ID length is then determined by the first level where the continuation probability falls below a threshold:
\begin{equation}
    L_{\mathrm{act}}
    =
    \min
    \left\{
    l:
    g_l \le \theta
    \right\},
\label{eq:activated_length}
\end{equation}
where $\theta$ is a stopping threshold. If no level satisfies the stopping condition, the tokenizer uses the maximum allowed length $L_{\max}$.

We train the variable-length tokenizer with a reconstruction objective and a length distribution regularizer:
\begin{equation}
    \mathcal{L}
    =
    \mathcal{L}_{\mathrm{rec}}^{\mathrm{act}}
    +
    \lambda
    \mathrm{KL}
    \left(
    P_{\mathrm{len}}
    \,\|\, Q
    \right),
\label{eq:varlen_sid_loss}
\end{equation}
where $\mathcal{L}_{\mathrm{rec}}^{\mathrm{act}}$ is the reconstruction loss computed using only the activated residual levels, and $P_{\mathrm{len}}$ is the empirical distribution of activated code lengths induced by ALGN. The prior $Q$ controls the expected length distribution. In practice, we use a decaying prior such as the geometric distribution
\begin{equation}
    Q(l)
    =
    (1-\gamma)^{l-1}\gamma .
\label{eq:length_prior}
\end{equation}
This regularizer prevents the tokenizer from trivially activating all residual levels and encourages a compact length distribution. Compared with directly penalizing the length of each sample, a distributional prior is more flexible because it can incorporate prior knowledge about the long-tail nature of user behaviors.

Overall, ALGN implements a local marginal utility rule for behavioral tokenization. Fixed-length RQ-VAE assigns the same capacity to every behavioral period, while ALGN dynamically allocates capacity according to the remaining residual information, quantization uncertainty, and marginal code cost. Therefore, high-information periods receive deeper semantic IDs, while routine or redundant periods stop early. This converts the Behavioral Densing Law from a scale-level capacity principle into an instance-level adaptive tokenization mechanism.

\subsection{Experimental Results}
\label{subsec:varlen_experiment}

We first compare downstream representation quality and the utilized SID capacity of each method. 
As studied in Section \ref{sec:densing}, all classification and retrieval tasks exhibits the similar performance trend thus only classification metrics on paybill data are reported.
The performance comparison is conducted using dataset size at 180 days sequence length and 0.1B user count, with the input setting held constant across all methods.
We set the length prior $Q$ in Eq.~\ref{eq:varlen_sid_loss} as a geometric distribution with $\gamma=0.3$ and compare ALGN with four baselines: fixed-length RQ-VAE, a prior-based heuristic method, a residual-norm-only variant, and an information-entropy-only variant. This comparison evaluates whether joint gating with residual information and semantic uncertainty provides better capacity allocation than fixed, prior-only, or single-signal allocation.


\subsubsection{Comparison}

\begin{table}[!ht]
    \caption{Comparison of different tokenization methods. SID capacity
    usage is used to measure efficiency, with classification metrics reported for performance evaluation.}
    \label{tab:algn_overall_results}
    \centering
    \resizebox{0.99\linewidth}{!}{
        \begin{tabular}{c|c|ccc}
        \toprule
        \textbf{Method}
        &
        \textbf{Capacity (\%)$\downarrow$}
        &
        \textbf{AUC (\%)$\uparrow$}
        &
        \textbf{KS (\%)$\uparrow$}
        &
        \textbf{Acc (\%)$\uparrow$}
        \\
        \midrule

        RQ-VAE\citep{lee2022autoregressive}
        &
        100.00
        &
        74.56
        &
        39.02
        &
        82.44
        \\

        VQ-VAE\citep{vqvae}
        &
        316.23
        &
        73.45
        &
        37.95
        &
        82.47
        \\

        RQ-Kmeans\citep{rqkmeans}
        &
        100.00
        &
        73.77
        &
        38.34
        &
        82.85
        \\

        SARQ\citep{sarq}
        &
        76.24
        &
        75.36
        &
        41.28
        &
        83.16
        \\

        \textbf{ALGN}
        &
        \textbf{63.47}
        &
        \textbf{76.43}
        &
        \textbf{43.31}
        &
        \textbf{83.52}
        \\

        \bottomrule
        \end{tabular}
    }
\end{table}

Results in Table~\ref{tab:algn_overall_results} shows that ALGN surpasses the best baseline SARQ by increasing 1.07\%/2.03\%/0.36\% AUC/KS/Acc and saving 13.23\% capacity.


The comparison among variable-length variants further supports the design of ALGN. The heuristic method slightly reduces SID usage but brings only marginal performance gains, suggesting that prior length statistics alone are insufficient for instance-level capacity allocation. Residual norm and information entropy each provide stronger improvements, indicating that both remaining reconstruction information and semantic uncertainty are useful signals for estimating the marginal utility of additional code levels. ALGN achieves the best performance and the lowest SID usage by jointly modeling residual information, code uncertainty, and marginal code cost.

These results provide an instance-level validation of the Behavioral Densing Law. Instead of assigning a fixed capacity to all behavioral periods, ALGN allocates more residual codes only when additional capacity is likely to bring useful information. As a result, high-information periods receive deeper semantic IDs, while routine or redundant periods stop early. This improves downstream representation quality while reducing average tokenization cost.

\subsubsection{Ablation Study}

\begin{table*}[!ht]
    \caption{Ablation study on each component of ALGN.}
    \label{tab:algn_ablation}
    \centering
    \resizebox{0.99\linewidth}{!}{
        \begin{tabular}{c|c|ccc}
        \toprule
        \textbf{Method}
        &
        \textbf{Capacity(\%) $\downarrow$}
        &
        \textbf{AUC (\%)$\uparrow$}
        &
        \textbf{KS (\%)$\uparrow$}
        &
        \textbf{Acc (\%)$\uparrow$}
        \\
        \midrule

        RQ-VAE\citep{lee2022autoregressive}
        &
        100.00
        &
        74.56
        &
        39.02
        &
        82.44
        \\

        Heuristic Variable-Length Baseline
        &
        86.91
        &
        74.43
        &
        38.10
        &
        82.39
        \\
        
        \hdashline

        w/o Uncertainty Signal
        &
        77.83
        &
        75.11
        &
        40.21
        &
        82.77
        \\

        w/o Residual Signal
        &
        76.24
        &
        75.36
        &
        41.28
        &
        83.16
        \\

        \hdashline

        w/o Regularization
        &
        73.42
        &
        76.01
        &
        42.21
        &
        83.35
        \\

        w/ SID Length Regularization
        &
        \textbf{63.16}
        &
        76.22
        &
        42.99
        &
        83.47
        \\

        w/ Random Distribution Regularization
        &
        70.15
        &
        76.16
        &
        42.88
        &
        83.45
        \\

        \textbf{ALGN}
        &
        63.47
        &
        \textbf{76.43}
        &
        \textbf{43.31}
        &
        \textbf{83.52}
        \\

        \bottomrule
        \end{tabular}
    }
\end{table*}

Compared with the heuristic variable-length baseline, ALGN reduces capacity usage from 86.91\% to 63.47\%, while improving AUC/KS/Acc at 2.00\%/5.21\%/1.14\% respectively with the adaptive variable SID length control.

Removing either adaptive signal degrades efficiency and performance. Without the uncertainty signal, capacity usage increases to 77.83\%, while AUC/KS/Acc decrease by 1.32/3.10/0.75\% relative to ALGN. Removing the residual signal yields similar degradation.

Regularization ablation further validates the proposed design. Removing regularization increases capacity usage to 73.42\% and decreases AUC/KS/Acc to 76.01\%/42.21\%/83.35\%. Although direct SID-length regularization achieves slightly lower capacity usage, ALGN improves AUC/KS/Acc by 0.21/0.32/0.05\%, demonstrating the best overall efficiency–performance trade-off with the incorporation of prior knowledge about the long-tail user behavior nature. It also outperforms random-distribution regularization by 0.27/0.43/0.07 \% while reducing capacity usage by 6.68\%.

\paragraph{Effect on Densing Law.}

\begin{figure*}[!ht]
    \centering
    \includegraphics[width=0.7\linewidth]{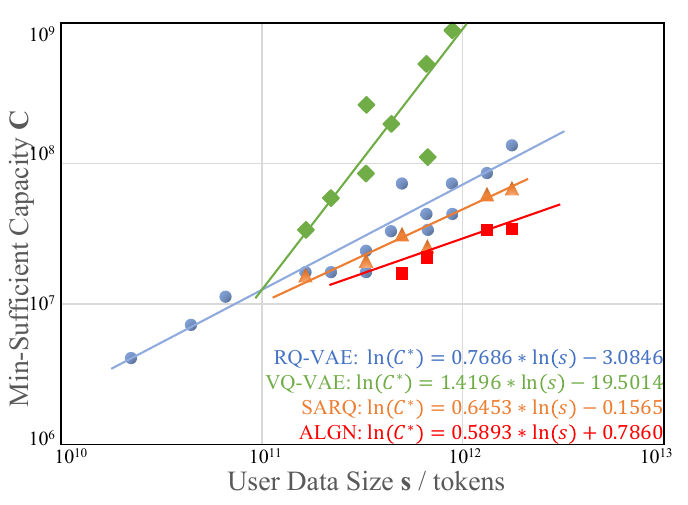}
    \caption{
    Effect on densing law for ALGN under classification task.}
    \label{fig:algn_densing_law}
\end{figure*}

We also conduct experiments to evaluate the ALGN's effect on our proposed densing law. Results in Figure \ref{fig:algn_densing_law} show that our method effectively decreases the scaling slope to $\sim$0.59, 
probably by reducing representation space redundancy $\mathcal{E}_{\phi}$.

\section{Conclusion \& Limitation}
\label{sec:conclusion}

The belief that more data and larger models always produce better user representations has underpinned years of industrial investment in behavioral sequence modeling. This work presents the first systematic, large-scale empirical test of that assumption. On Alipay's production data spanning hundreds of millions of users, we demonstrate that all three conventional scaling axes, user population, temporal horizon, and model capacity, exhibit clear saturation thresholds. Increasing users beyond 0.03B or temporal windows beyond 60 days delivers rapidly diminishing returns. More strikingly, scaling model parameters past 0.2B under fixed data continues to lower training loss but fails entirely to improve downstream accuracy, revealing that optimization objective and representation quality are not the same thing. These three findings collectively define the \textbf{scaling walls} for industrial user behavioral modeling. They converge on a single diagnosis: the bottleneck is not the volume of data, but the predictive information carried by each behavioral token.

This finding sets up our central proposal: a shift from data scaling to \emph{density scaling}. We introduce the concept of \emph{data densing}, drawing an explicit parallel to the densing principle at the model level that has driven progress in compact language model design. Through RQ-VAE tokenization and a matched-budget evaluation protocol, we operationalize this principle: compact, fixed-length discrete codes carry more task-relevant signal than the original raw sequences. The densified representation shifts the scaling curve to a more favorable exponent, enabling smaller encoders operating on shorter inputs to outperform larger models trained on longer raw histories.

For practitioners, our findings translate into concrete, deployable guidelines. The saturation thresholds reported here can directly inform data retention policies, training budget allocation, and model architecture decisions in production environments. The discrete tokens produced by RQ-VAE tokenization are inherently task-agnostic, reusable across user targeting, profile prediction, CTR estimation, and beyond, amortizing the one-time cost of tokenizer training across an entire application portfolio. At a time when the computational and environmental costs of indiscriminate data scaling are drawing increasing scrutiny, density scaling offers a principled alternative: better representations from the data we already have, not more data collected at ever-increasing cost.

We acknowledge a limitation of the current study: our experiments focus on a single data modality, and the generalizability of the Densing Law to other behavioral modalities, such as video consumption, remains to be validated. We are also further evaluating the proposed law on public benchmark datasets to assess its robustness beyond data collected from a single platform.

These limitations point toward a broader research agenda. We believe that \emph{information density} should be treated as a first-class design objective in large-scale user modeling, alongside model capacity and data volume. The implications extend beyond this work: if density can be measured and optimized for, then the practical question shifts from ``how much data can we collect?'' to ``how do we design representations that maximize signal per token?''. We invite the community to explore adaptive tokenization strategies, multi-modal density estimation, and the theoretical foundations of data densing, as a collective step \textbf{from scaling walls to densing gains}: toward representation systems that scale not by consuming more, but by extracting more from what they already have.

\bibliographystyle{plainnat}
\bibliography{main}

\section{Appendix}
\label{sec:appendix}

\subsection{Training Loss Convergence across Temporal Horizons}
\label{subsec:appendix_loss}

Training loss during pretraining is another signal for scaling comparison, which reflects the optimization behavior of the pretraining objective. This loss is computed according to Eq.~\ref{eq:contra_loss}.

\begin{figure}[htbp]
  \centering
  \includegraphics[width=\linewidth]{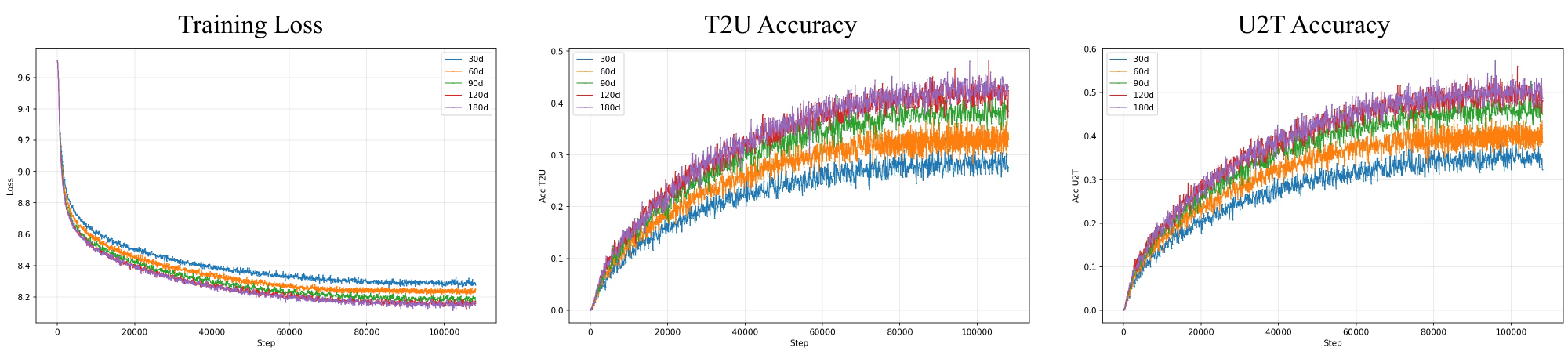}
  \caption{Contrastive alignment loss over $100{,}000$ training steps for models pre-trained on behavioral sequences spanning $D \in \{30, 60, 90, 120, 180\}$ days. While all configurations converge, the reduction in final loss diminishes sharply as $D$ increases. The curves for $D \ge 90$ days nearly overlap, indicating that additional historical days contribute minimal new optimization signal, consistent with the downstream metric saturation observed in Section~\ref{sec:scaling_analysis}.}
  \label{fig:scaling_law_loss}
\end{figure}

It can be observed in Figure \ref{fig:scaling_law_loss} As the behavioral timespan scales from 30 to 180 days, the training loss converges to progressively lower values, decreasing from approximately 8.28 to 8.15. Meanwhile, the contrastive-learning accuracies exhibit a consistent positive scaling trend: T2U accuracy increases from approximately 0.29 to 0.42, while U2T accuracy rises from approximately 0.35 to 0.50, indicating that larger-scale behavioral inputs improve representation alignment in both directions.

\subsection{Generalization across different data sources}
\label{subsec:appendix_loss}

\begin{figure}[htbp]
  \centering
  \includegraphics[width=\linewidth]{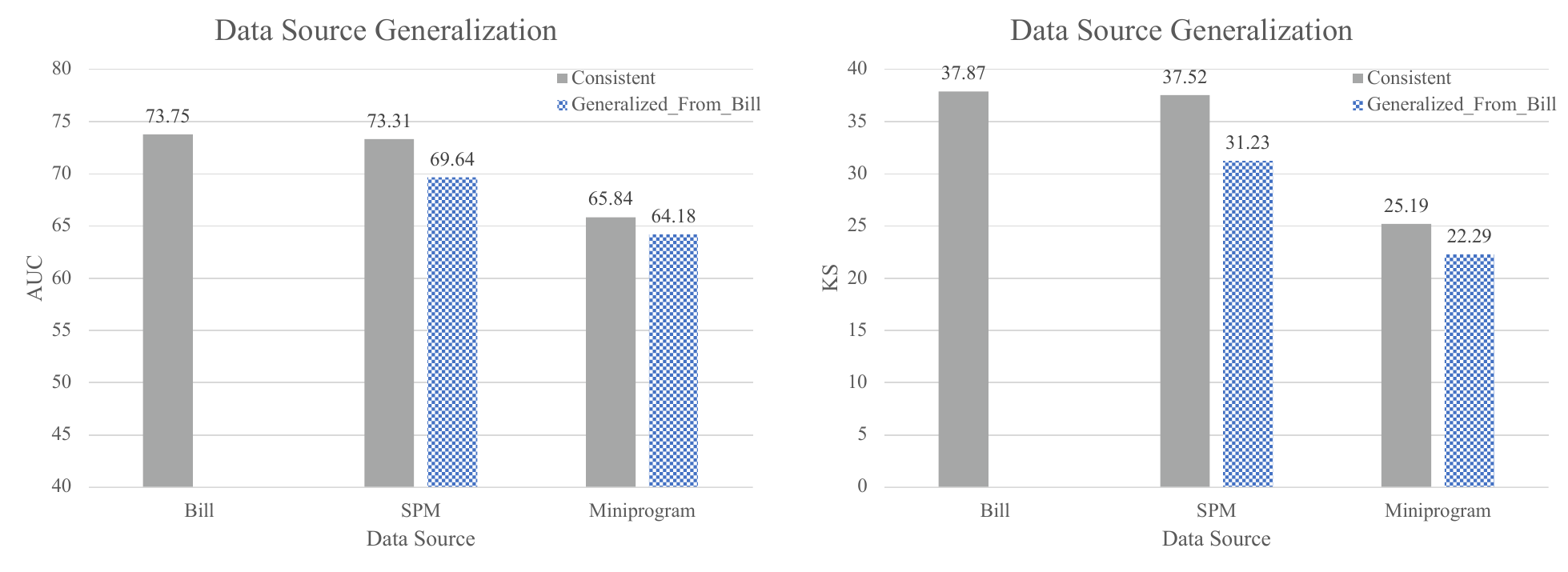}
  \caption{Data source generalization.}
  \label{fig:data_source_generalization}
\end{figure}

As shown in Figure \ref{fig:data_source_generalization} , the compression strategy exhibits strong generalizability and adaptability across heterogeneous data sources. When the compression configuration derived from the Bill dataset is directly transferred to SPM and Miniprogram, it preserves approximately 95.0\% and 97.5\% of the AUC achieved under the source-consistent setting, respectively. A similar trend is observed for KS, with retention rates of 83.2\% on SPM and 88.5\% on Miniprogram. These results indicate that the compression strategy is not strongly dependent on source-specific data characteristics and can maintain competitive predictive performance under cross-source distribution shifts, demonstrating its robustness and practical applicability across diverse data environments.






\end{document}